\documentclass{iopart}

\usepackage{iopams}
\usepackage{graphicx}
\usepackage{float}
\usepackage{xcolor}

\usepackage{hyperref}

\newcommand{\eqref}[1]{(\ref{#1})}

\usepackage{bm} 

\newcommand{\arctanh}{{\rm arctanh}}

\newcommand{\dst}[0]{\ensuremath{\mathrm{d}}}
\newcommand{\mdis}[0]{\ensuremath{\mathcal{M}}}

\begin{document}

\title[Dynamical phase transition in the activity-biased fully-connected RFIM]
{Dynamical phase transition in the activity-biased fully-connected random field Ising model: connection with glass-forming systems}
\author{Jules Guioth$^{1,2}$, Robert L. Jack$^{1, 3}$}
\address{$^1$ Department of Applied Mathematics and Theoretical Physics, University  of Cambridge, Wilberforce Road, Cambridge CB3 0WA, United Kingdom}
\address{$^2$ Univ. Lyon, \'{E}NS de Lyon, Univ. Claude Bernard, CNRS, Laboratoire de Physique, F-69342 Lyon, France}
\address{$^3$ Yusuf Hamied Department of Chemistry, University of Cambridge, Lensfield Road, Cambridge CB2 1EW, United Kingdom}
\ead{rlj22@cam.ac.uk}


\begin{abstract}
We analyse biased ensembles of trajectories for the  random-field Ising model on a fully-connected lattice, which is described exactly by mean-field theory.  By coupling the activity of the system to a dynamical biasing field, we find a range of dynamical phase transitions, including spontaneous symmetry breaking into ordered states.  For weak bias, the phase behaviour is controlled by extrema of the free energy, which may be local minima or saddle points.  For large bias, the system tends to states of extremal activity, which may differ strongly from free energy minima.  We discuss connections of these results to random first-order transition theory of glasses, which motivates an extension of the analysis to random-field Ising models where the dynamical activity is not symmetric under magnetisation reversal.  
\end{abstract}


\section{Introduction}
\label{sec:glass}

This article analyzes a class of dynamical fluctuations in the random-field Ising model, motivated by the connection of this model to glass-forming systems~\cite{biroli2014random,biroli2018random1,biroli2018random2,franz2013glassy,franz2013universality}.
This connection is introduced first, followed by a description of the model and the particular fluctuations of interest.

\subsection{Static and dynamic phase transitions in theories of the glass transition}

Glass transitions occur in a broad range of systems, including structural and spin glasses \cite{berthier2011theoretical}. When approaching the glass transition, dynamics slow down dramatically while structure remains disordered.
It remains an open theoretical question as to whether the glass is a genuine phase (distinct from the liquid state by a sharp phase transition) or instead a crossover  \cite{berthier2011theoretical, arceri2020glasses}.

Two contrasting theories~\cite{biroli2013} both explain the slow dynamics near the glass transition via the proximity to some kind of phase transition.
The random first order transition (RFOT) theory describes a free energy landscape that changes in character as temperature decreases. At mean-field level~\cite{castellani2005spin}, the free energy landscape of a liquid at high temperature has a simple minimum that corresponds to a fluid state. When temperature decreases below a certain threshold a huge number of metastable states appear and the free energy landscape becomes rough, before the equilibrium ensemble is eventually dominated by low free energy minima which correspond to the glass state.  
This corresponds to a thermodynamic phase transition where replica symmetry is broken. 
The order parameter can be understood as the overlap between two copies of the system.  Phase transitions also appear if these systems are biased by a field (conventionally denoted by $\epsilon$) that is conjugate to the overlap~\cite{franz1997phase}.  The universal properties of these phase transitions are related to the random-field Ising model (RFIM) universality class~\cite{biroli2014random,biroli2018random1,biroli2018random2,franz2013glassy,franz2013universality}.

Another approach is that of dynamical facilitation, 
which is a dynamic point of view where the glass state appears as an inactive phase in trajectory space \cite{Chandler2010}. In this perspective, the glass transition is related to a sharp dynamical ``space-time'' phase transition between an active liquid state and an inactive glass state.
Such dynamical phase transitions are characterised within a large deviation formalism, based on ensembles of trajectories \cite{garrahan2007dynamical, garrahan2009first, hedges2009dynamic}.
The existence of such dynamical phase transitions have been demonstrated in several different glassy systems, including kinetically constrained models~\cite{garrahan2007dynamical, garrahan2009first,Bodineau2012jsp,nemoto2014finite}, spin models~\cite{Jack2010rom, van2010second, jack2016phase, Turner2015}, and atomistic liquids~\cite{hedges2009dynamic,Malins2012-sens}; there are also relevant experiments~\cite{pinch17,Abou2018}.   They appear when systems' dynamical activity is biased by a field that is conventionally denoted by $s$.

\subsection{Static and dynamic phases of the RFIM: connection to RFOT theory}

Some glassy models can support both thermodynamic transitions and dynamical ones~\cite{Turner2015,jack2016phase}, induced by biasing fields $\epsilon$ and $s$ respectively.  This work shows that this situation also occurs in the RFIM.
The resulting dynamical picture has a rich structure, even for the mean-field (fully-connected) version of the model.

\begin{figure}
  \centering
  \includegraphics[width=0.8\linewidth]{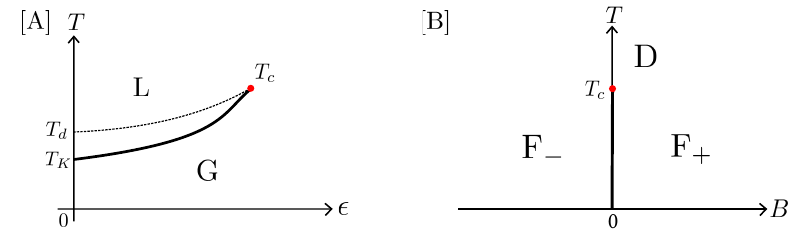}
  \caption{Sketches of the equilibrium  phase diagrams of the replica-biased glass-forming system [\textbf{A}] and of the (weak disorder) RFIM [\textbf{B}]. [\textbf{A}]: L is referring to the liquid phase and G the glass phase. The dashed line is the spinodal curve (the spinodal region thus corresponds to the region between the plain black line (binodal or first order transition line) and the dashed line). [\textbf{B}]: the notation D and $\mathrm{F}_{\pm}$ refers respectively to the disordered phase and the symmetric ferromagnetic phases (with positive and negative magnetisation). 
 In both figures, the heavy black lines are first-order transition lines that end in second order critical points (with $T=T_{c}$).
  The RFOT first-order transition line $T_{\rm eq}(\epsilon)$ has a non-trivial dependence on $\epsilon$ whereas the (weak disorder) RFIM one is vertical, located at $B=0$ for $0 \leqslant T \leqslant T_{c}$.}
  \label{fig:0}
\end{figure}

Our results will be compared with 
a corresponding analysis for the mean-field Ising model~\cite{GuiothJack2020}.  
Moreover, since the RFIM is related to the RFOT theory, the results also have implications in the glassy context.
To explain that connection, we briefly 
review the correspondence between the equilibrium phase diagrams of RFOT glasses and the RFIM.
 (For the RFIM, we consider the case of weak or moderate disorder, see below.)
 
The equilibrium phase diagrams are sketched in
Fig.~\ref{fig:0}. 
Both cases include first-order transition lines that end in critical points, whose universal properties are the same~\cite{biroli2014random,biroli2018random1,biroli2018random2,franz2013glassy,franz2013universality}.
In the RFIM, the first-order transition line is at zero magnetic field ($B=0$), and the critical point involves spontaneous breaking of the (global) spin-flip symmetry.   
In contrast, the transition line of the RFOT theory does not follow a line of symmetry, we describe it by a function $\epsilon_{\rm eq}(T)$.

The natural analogy between RFIM and RFOT is to identify the magnetisation $m$ of the RFIM with the order parameter $q$ (overlap) of RFOT. The conjugate fields for these order parameters are the magnetic field $B$ of the RFIM, and the bias $\epsilon$ : in fact the natural RFOT analogue of $B$ is the difference $\epsilon-\epsilon_{\rm eq}(T)$ between the bias and its value on the phase transition line.  
Finally, we recall that the disorder in the RFIM originates in a quenched random field (which will be called $h_i$ in the following); the corresponding RFOT object is the quenched random configuration that appears in the definition of the overlap.   The relationships between these parameters is summarized in Table~\ref{tab:translation_table}.

In addition to these parameters, the RFIM behaviour also depends on the strength of the random field, here denoted by $\eta$.
In the RFOT picture, the analogue of this quantity is related in a non-trivial way to the structure of the liquid~\cite{biroli2014random}.  The most interesting case for glasses corresponds to the RFIM behaviour for weak or moderate disorder, as shown in Fig.~\ref{fig:0}.  For stronger disorder, the phase transition disappears, and there is no ferromagnet.

As advertised above, this work analyses the RFIM where one additionally biases the dynamics by a field $s$.  
The resulting situation depends on the temperature, the field $B$, the random field strength $\eta$, and on $s$.   
One may imagine extending the RFIM phase diagram of Fig.~\ref{fig:0} by adding a new axis for the bias $s$.  
We note that theoretical connection between the RFOT theory to the RFIM does not extend to dynamical properties.  Still, it is interesting to compare the effects of the bias $s$ on the RFIM and compare with its effects on glassy systems. 

With these points in mind, the paper is organised as follows. In section \ref{sec:th}, we briefly introduce the version of the RFIM that we consider, and the corresponding ensembles of trajectories. In section \ref{sec:eq}, we recap the equilibrium phase diagram of the RFIM for two different disorder distributions, the Gaussian and bimodal cases. 
The core of the analysis is provided in section \ref{sec:s_ens_RFIM} where we detail the phase diagrams of the activity-biased mean-field RFIM. 
Section \ref{sec:asym-mob} explains how these results change when one considers systems without global spin-flip symmetries, which is relevant for the connection to glassy systems.  Conclusions are summarised in Sec.~\ref{sec:conc}.

\section{Theory: activity-biased mean-field RFIM}\label{sec:th}

\begin{table}
\raggedleft
  \begin{tabular}{|c|c|}
    \hline
    Glass-forming systems (RFOT) & Magnetic system (RFIM) \\
    \hline \hline
    Overlap $q \in [0,1)$                  & Magnetisation $m\in[-1,1]$ \\
    \hline
    Tilt $\epsilon-\epsilon_{\rm eq}(T)$              & Magnetic field $B$ \\
    \hline
    Disorder $h_{\mathcal{C}_{\rm ref}}(x)$ & Random field $h(x)$   \\
    \hline
  \end{tabular}
  \caption{Translation table between replica-biased glass-forming system as analysed by RFOT and magnetic systems.}
  \label{tab:translation_table}
\end{table}

\subsection{Model definition and biased ensembles}\label{sec:model_def}

\subsubsection{Static properties.}

We consider the RFIM on a fully-connected lattice, whose properties can be computed by a version of Curie-Weiss theory.  
It consists of $N$ spins, the 
$i$th spin is $\sigma_{i}=\pm 1$, and the overall configuration is $\bm{\sigma} = {\{ \sigma_{i} \}}_{i=1}^{N}$.  
In addition to their exchange interaction, each spin interacts with an external \emph{quenched} random magnetic field $\bm{h}={\{ h_{i} \}}_{i=1}^{N}$. 
The random field $h_{i}$ is assumed to have mean zero, and a variance of order $1$. The parameter $\eta$ is the coupling to the disorder, so
the energy of the mean-field RFIM is 
\begin{equation}
  \label{eq:1a}
  E(\bm{\sigma}|\bm{h}) = -  \frac{1}{N} \sum_{ij} \sigma_i \sigma_j  - B \sum_i \sigma_i - \eta \sum_i h_i \sigma_i  \, ,
\end{equation}
where $B$ is an external magnetic field.  In the exchange term, the coupling constant (sometimes denoted $J$) has been fixed at unity, so every pair of spins interacts with a coupling $1/N$.   This does not lose any generality, it simply means that all energies are measured in units of the coupling constant.

The magnetisation is
\begin{equation}
  \label{eq:4}
  m = \frac{1}{N} \sum_{i=1}^{N} \sigma_{i}
\end{equation}
and the (intensive) overlap between the system's configuration $\bm{\sigma}$ and the random field $\bm{h}$ is
 \begin{equation}
  \label{eq:5}
  \tau = \frac{1}{N} \sum_{i=1}^{N} \sigma_{i} h_{i}  \; .
\end{equation}
Hence
\begin{equation}
  \label{eq:1b}
  E(\bm{\sigma}|\bm{h}) = - N \left( m^{2} + B m + \eta \tau \right) \, .
\end{equation}
The Boltzmann equilibrium distribution (for a given realisation of the random field) is
\begin{equation}
  \label{eq:6}
  P_{\rm eq}(\bm{\sigma}|\bm{h}) = \frac{1}{Z_{\rm eq}(\beta | \bm{h})} e^{-\beta E(\bm{\sigma}|\bm{h})} \, ,
\end{equation}
with $Z_{\rm eq}(\beta | \bm{h})$ the canonical partition function. 

The random field $\bm{h}$ will be averaged over a distribution $P_{\rm dis}(\bm{h})$ in which the $h_i$ are independently and identically distributed,
$P_{\rm dis}(\bm{h}) = \prod_{i=1}^{N} p_{\rm dis}(h_{i})$.  For the single site probability density $p_{\rm dis}$, we consider two common distributions: a Gaussian distribution $p_{\rm G}$ \cite{schneider1977random} and a discrete bimodal density distribution $p_{\rm bi}$ \cite{luttinger1976exactly,aharony1978tricritical}:
\begin{eqnarray}
  \label{eq:3:1}
  p_{\rm G}(h) &= & \frac{1}{\sqrt{2\pi}} e^{-h^{2}/{2}} \, \\
  \label{eq:3:2}
  p_{\rm bi}(h) & = & \frac{1}{2}\delta(h-1) + \frac{1}{2}\delta(h+1) \, ,
\end{eqnarray}
where $\delta$ indicates a Dirac delta function.

The free energy of the RFIM is related to the partition function as $(-1/\beta) \ln Z_{\rm eq}(\beta | \bm{h})$.  Averaging the disorder and taking thermodynamic limit gives the free energy per spin
\begin{equation}
\label{equ:bar-f-eq}
\bar{f}_{\rm eq} = -\lim_{N\to\infty} \frac{1}{N\beta} \overline{ \ln Z_{\rm eq}(\beta | \bm{h}) }
\end{equation}
where $\overline{(\cdot)} = \int (\cdot) P_{\rm dis}(\bm{h}) {\rm d}\bm{h}$  indicates the disorder average.

\subsubsection{Dynamics}\label{sec:theory:dyn}

We consider Markov jump dynamics in continuous time. Each jump involves a single spin changing its state.  The jump rates obey detailed balance with respect to $P_{\rm eq}$, so the transition rate from state $\bm{\sigma}$ to $\bm{\sigma}'$ takes the form
\begin{equation}
  \label{eq:def:general_transition_rates}
  w(\bm{\sigma}'| \bm{\sigma}) = \widetilde{\chi}(\bm{\sigma}, \bm{\sigma}') 
  \exp\left[ - \frac{\beta}{2} \Delta E(\bm{\sigma}, \bm{\sigma}') \,  \right]
\end{equation}
where the function $\widetilde{\chi}$ is symmetric, that is $\widetilde{\chi}(\bm{\sigma}, \bm{\sigma}')= \widetilde{\chi}(\bm{\sigma}', \bm{\sigma})$ and $\Delta E(\bm{\sigma}, \bm{\sigma}') = E(\bm{\sigma}')-E(\bm{\sigma})$.  The quantity $\widetilde{\chi}$ is interpreted as a mobility \cite{kaiser2018canonical,maes2007and}.  We consider the general class of mobilities 
\begin{equation}
  \label{eq:12}
  \widetilde{\chi}(\bm{\sigma},\bm{\sigma}') = \chi\left(\frac{\beta}{2} \Delta E(\bm{\sigma}, \bm{\sigma}') \right) \, ,
\end{equation}
with $\chi$ an even function. We focus on the Glauber dynamic rule, which is $\chi(x) = [\cosh(x)]^{-1}$.
Define also the escape rate from state $\bm{\sigma}$ as
\begin{equation}
  \label{eq:escrate}
  r(\bm{\sigma}) = \sum_{\bm{\sigma}' (\neq \bm{\sigma})}w(\bm{\sigma}'|\bm{\sigma}) \; .
\end{equation}
Given a system in state $\bm{\sigma}$, the time until the next spin flip is exponentially distributed with mean $r(\bm{\sigma})^{-1}$.

\subsection{Conditioned and biased trajectory ensembles on activity}
\label{sec:biased_ens_ld}

As discussed in Sec.~\ref{sec:glass}, analysing large fluctuations of the dynamical activity can reveal interesting behavior in glassy systems, and in other models too.  This Section describes the framework for such analysis, see also~\cite{lecomte2007thermodynamic,derrida2007non,garrahan2009first,chetrite2015nonequilibrium,Jack2019}.

Let $\Theta_{[0,T]} = \{ \bm{\sigma}(t) \}_{t \in [0,T]}$ denote a trajectory of the system on the time interval $[0,T]$.
Also define $\mathcal{K}[\Theta_{[0,T]}]$ as the total number of jumps in the trajectory $\Theta_{[0,T]}$.  (That is, the total number of configuration changes or ``spin flips'' in the time interval $[0,T]$.)
Several different observables have been used to quantify dynamical activity. 
The formalism of this section is general: we write ${\cal A}$ for a generic (time-averaged) measure of activity, but our analysis of the RFIM is restricted to
\begin{equation}
  \label{eq:7}
  \mathcal{A}[\Theta_{[0,T]}] = \frac{1}{T}\mathcal{K}[\Theta_{[0,T]}]
\end{equation}
so that the activity is the time-averaged empirical jump rate.

Since the system is a finite Markov chain, the activity obeys a large-deviation principle: as $T\to\infty$ then
\begin{equation}
  \label{eq:LDP_time-integrated_activity}
  P\left(\mathcal{A}[\Theta_{[0,T]}] \approx a\right) {\sim} e^{-T I(a)} \, ,
\end{equation}
where $I(a)$ is the large-deviation function (or rate function).  Denote the typical value of ${\cal A}$ by $a^\ast$.  
At equilibrium, most of the observed trajectories have $\mathcal{A}\approx a^{\ast}$, 
so that
$I(a^{\ast}) = 0$ which is the minimal possible value of the rate function. Other values of $a$ involve large fluctuations whose probabilities are quantified by $I(a)$. 

In addition to the probability of such events, it is also possible to characterise their mechanism~-- that is, the behaviour of the (very unlikely) trajectories $\Theta_{[0,T]}$ that realise the non-typical activity $a$.  
This is achieved by the biased ensemble 
 -- sometimes called the $s$-ensemble.  Let $P( \Theta_{[0,T]} )$ be the probability density for trajectory $\Theta_{[0,T]} $ under the equilibrium dynamics and let $\langle \cdot \rangle$ be an average with respect to this dynamics.  Then the  probability of trajectory $\Theta_{[0,T]} $ in the biased ensemble is 
\begin{equation}
  \label{eq:biased_ensemble}
  P_{s}( \Theta_{[0,T]} ) = \frac{1}{Z(s,T)} P( \Theta_{[0,T]} ) e^{-s T \mathcal{A}[ \Theta_{[0,T]} ]} \; ,
\end{equation}
where the normalisation constant
\begin{equation}\label{eq:partition_function_biased_ens}
  Z_N(s,T|\bm{h}) = \left\langle \exp(-s T \mathcal{A}[\Theta_{[0,T]}]) \right\rangle  
\end{equation}
is similar to the partition function in equilibrium statistical mechanics.  Based on this analogy, we define a dynamical free energy
\begin{equation}  
  \label{eq:def:Psi}
  \Psi_N(s|\bm{h}) = \lim_{T\to\infty} \frac{1}{T} \log Z_N(s,T|\bm{h})
\end{equation}
which is analogous to the (negative of the) free energy in the canonical ensemble.  

Since the system is finite, $\Psi_N$ is an analytic function of $s$.  To analyse 
dynamical phase transitions, we average the disorder and take the thermodynamic limit
\begin{equation}
\psi(s) = \lim_{N\to\infty} \frac{1}{N} \overline{\Psi_N(s|\bm{h})} 
\label{eq:small-psi}
\end{equation}
(recall that the overbar indicates the disorder average).
The resulting function $\psi$ may have singularities, which correspond to dynamical phase transitions.
We note that the definition of $\psi$ requires two limits (of $N,T\to\infty$).  
As discussed in~\cite{Jack2019,JackNemoto2019}, one expects quite generally that the limits of large $N,T$ commute with each other, although other properties of the biased ensemble can depend strongly on the relative size of $N$ and $T$.

The free energy $\Psi$ can be characterised as the largest eigenvalue of 
a tilted operator $\mathcal{W}_{s}$ whose matrix elements are
\begin{equation}
  \label{eq:27}
  \left(\mathcal{W}_{ s}\right)_{\bm{\sigma}'\!,\bm{\sigma}} = e^{-s} w(\bm{\sigma}'|\bm{\sigma}) - r(\bm{\sigma})\delta_{\bm{\sigma},\bm{\sigma}'} \, ,
\end{equation}
where $\bm{\sigma},\bm{\sigma}'$ denote configurations, $w$ is the transition rate from \eqref{eq:def:general_transition_rates} and $r$ is the escape rate \eqref{eq:escrate}.
The matrix ${\cal W}_s$ can be symmetrised, and its largest eigenvalue obeys a variational principle \cite[Eq. (27)]{garrahan2009first}:
\begin{equation}
\fl 
  \label{eq:varppl}
  \Psi_N(s|\bm{h}) = {\max}_{\pi} \left\{ \sum_{\bm{\sigma}} \left[ - r(\bm{\sigma})\pi(\bm{\sigma}) + e^{-s} \sum_{\bm{\sigma}'(\neq\bm{\sigma})} \sqrt{w(\bm{\sigma}|\bm{\sigma}')w(\bm{\sigma}'|\bm{\sigma})
  \pi(\bm{\sigma})\pi(\bm{\sigma}')} \right]  \right\} 
\end{equation}
where the maximisation is over probability distributions $\pi$ for configurations, normalised as $\sum_{\bm{\sigma}} \pi(\bm{\sigma})=1$. 
It can be shown \cite{Jack2019,garrahan2009first} that the distribution $\pi_{s}$ that realises the maximum corresponds to the stationary probability distribution of the biased-ensemble. (To be precise, the distribution is stationary up to transient corrections for times $t$ close to the boundaries at $t=0$ and $t=T$.)
This variational formula will be used in the following to characterise the biased ensemble for the RFIM.

\section{Equilibrium behaviour of the mean-field RFIM}\label{sec:eq}

To provide context for our analysis of dynamical phase transitions, this Section reviews the equilibrium phase behaviour of the RFIM, emphasising the differences that can appear between different disorder distributions (Gaussian and bimodal here) \cite{schneider1977random, aharony1978tricritical, krapivsky2010kinetic}.   The analysis proceeds by minimizing a suitable Landau free energy, which serves as a warm-up for maximization of the dynamical free energy $\Psi$ in Sec.~\ref{sec:s_ens_RFIM}.

\subsection{Equilibrium free energy and stationary states}
\label{sec:eq_free_en}

\newcommand{\mm}{{\cal M}}
\newcommand{\pdis}{p_{\rm dis}}

The mean-field nature of the spin-spin interaction allows one to compute the equilibrium behaviour of the mean-field RFIM by gathering the spins according to their random fields.
As detailed in \ref{app:eq_landau_free_en}, the state of the system can be fully characterised in terms of a function $\mm$, such that $\mm(h)$ is the magnetisation of those spins whose random field $h_i=h$.  For example, in the case of discrete bimodal disorder then $\mm(h) = m_+ \delta(h-1) + m_- \delta(h+1)$ so the function $\mm$ is fully specified by two numbers, which are the magnetizations of the subsets of spins with $h_i=\pm h$.  For continuous disorder then $\mm$ is a non-trivial function of $h$. 

For large systems $(N\to\infty)$, the disorder-averaged equilibrium Landau free energy $\bar{f}$ may be expressed as a functional of ${\cal M}$ using \eqref{eq:38},
\begin{equation}
  \label{eq:49}
  \bar{f}[\mm] = - (m^{2} + Bm + \eta \tau) - \frac{1}{\beta} \int \pdis(h) S(\mm(h)) \dst h \, ,
\end{equation}
where the magnetisation and overlap are expressed in terms of $\mm$ via
\begin{eqnarray}
\label{eq:m-mm}
m & = \int \pdis(h) \mm(h) \dst h \\
\tau & =\int \pdis(h) h \mm(h) \dst h
\end{eqnarray}
and $S(\mm)$ is given by (\ref{eq:entropy}).

The free energy $\bar f$ is to be minimized over the distribution $\mm$.
The solutions are denoted by $\mm^{\ast}$, they satisfy $\delta \bar{f}/\delta \mm(h) = 0$. This yields a self-consistent equation
\begin{equation}
  \label{eq:37}
 \mm^{\ast}(h) = \tanh(2\beta m^{\ast} + \beta B + \beta \eta h) \, ,
\end{equation}
with $m^{\ast} = \int \pdis(h) \mm^{\ast}(h) \, \mathrm{d}h$.
The right hand side of \eqref{eq:37} depends on $\mm^{\ast}$ only through $m^{\ast}$ so it is useful to
average over $h$ and hence obtain a self-consistency equation for $m^{\ast}$ alone:
\begin{equation}
  \label{eq:40}
  m^{\ast} = \int \pdis(h) \tanh(2\beta m^{\ast} + \beta B + \beta \eta h) \, {\rm d}h \, .
\end{equation}
One also obtains the overlap of the equilibrium state
\begin{equation}
  \label{eq:2}
  \tau^{\ast} = \int \pdis(h) h \tanh(2\beta m^{\ast} + \beta B + \beta \eta h) \, \mathrm{d}h \, .
\end{equation}
Note that while Eq.\ \eqref{eq:40} is a non-trivial self-consistency condition for $m^\ast$,  Eq.~\eqref{eq:2} is a simple formula which allows $\tau^\ast$ to be computed from $m^\ast$.

Finally, inserting \eqref{eq:37} into \eqref{eq:49}, the equilibrium free energy (\ref{equ:bar-f-eq}) may be obtained as
\begin{equation}
  \label{eq:39}
  \fl \qquad
  \bar{f}_{\rm eq} 
= 
\min_{m^{\ast}} \left\{ (m^{\ast})^{2} - \int \pdis(h)\ln\left[ 2 \cosh(2\beta m^{\ast} + \beta B + \beta \eta h) \right] \mathrm{d}h \right\} \, ,
\end{equation}
where the minimisation is over values of 
$m^{\ast}$ that obey (\ref{eq:40}), which correspond to extrema of (\ref{eq:49}).
The equilibrium magnetisation of the system is the $m^{\ast}$ that minimises $\bar{f}_{\rm eq}$.  (If the minimum is degenerate then the system is at a point of phase coexistence.)

\begin{figure}
  \centering
  \includegraphics[width=0.9\linewidth]{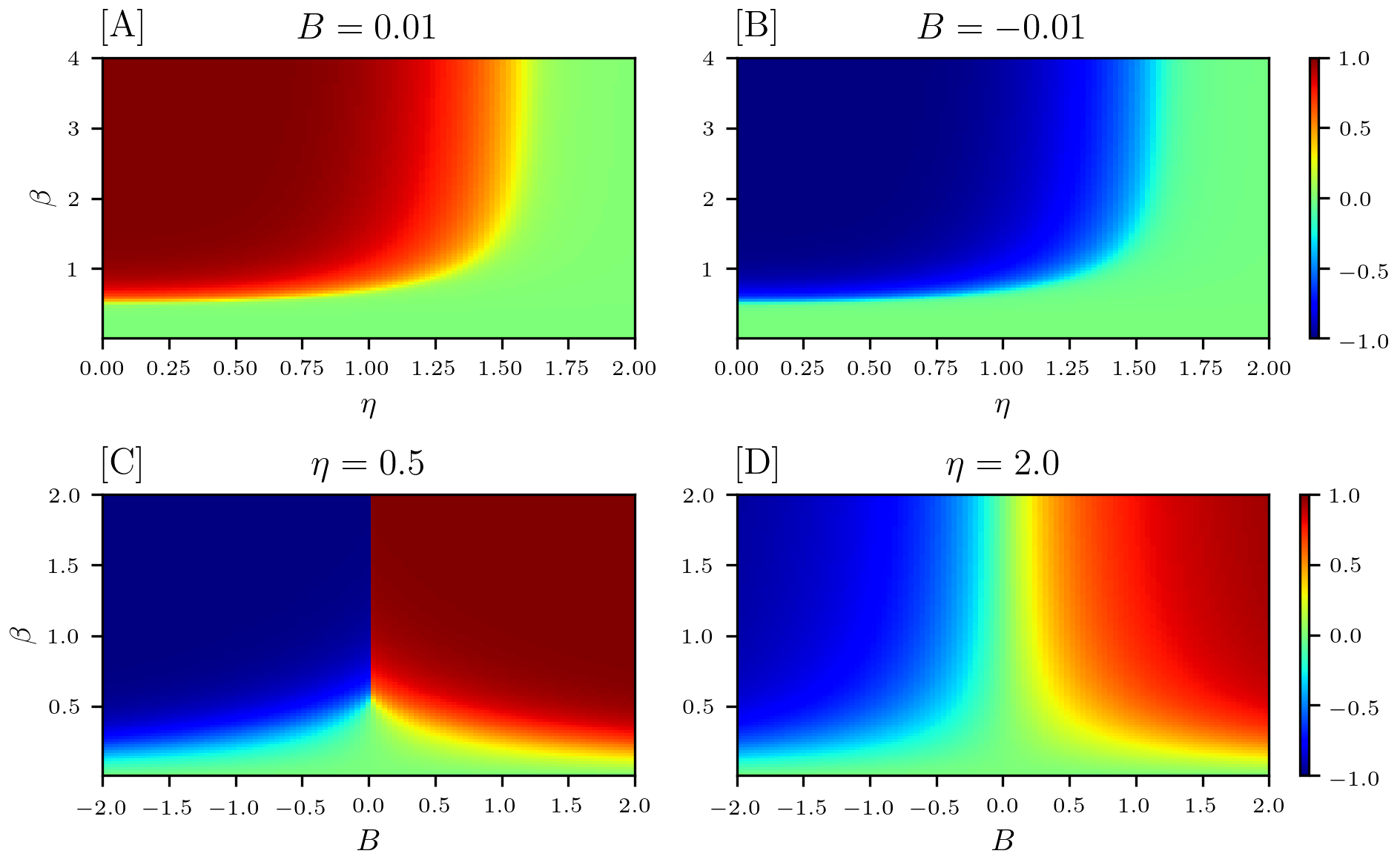}
  \caption{\textbf{Equilibrium phase diagram of the mean-field RFIM with Gaussian disorder.} 
  The colors indicate the global magnetisation $m^{\ast}$ (averaged over the disorder).
  \textbf{Panels A and B} show the $(\eta, \beta)$ plane for $B=\pm0.01$, the signature of a ferromagnetic state is that $m^{\ast}$ is discontinuous at $B=0$, which is signalled by a sharp contrast between the panels.  The transitions between ferromagnetic and paramagnetic states are continuous throughout the $(\eta,\beta)$ plane, including for large $\beta$.  \textbf{Panels C and D} show sections of the phase diagram in the $(B,\beta)$ plane for $\eta=0.5$ and $\eta=2.0$.  In panel C then $m^{\ast}$ is indeed discontinuous at $B=0$, for sufficiently large $\beta$ (ferromagnetic phase).  In panel D then $m^{\ast}$ is continuous everywhere, the ferromagnetic phase is destroyed by the disorder.}
    \label{fig:1}
\end{figure}

\subsection{Equilibrium phase diagrams}
\label{sec:eq-phase}

The equilibrium mean-field RFIM has already been analysed in \cite{schneider1977random} for Gaussian disorder, and in \cite{luttinger1976exactly, aharony1978tricritical} for bimodal disorder.
We show how these results can be obtained
by solving (numerically) the variational problem defined by~(\ref{eq:40}, \ref{eq:39}) to obtain $m^{\ast}$ as a function of $(\beta,\eta,B)$. 
This yields equilibrium phase diagrams, which are compared later with the dynamical phases obtained in the presence of the biasing field $s$.

\subsubsection{Gaussian disorder.}
\label{sec:eq_gaussian_dis}

Fig.~\ref{fig:1} shows the equilibrium phase diagram of the mean-field RFIM with Gaussian disorder, which is the distribution most commonly considered in the literature~\cite{schneider1977random,krapivsky2010kinetic}. 
In the case without disorder ($\eta=0$), we recover the usual second-order phase transition of the mean-field Ising model at critical temperature {$\beta=\beta^*_0=0.5$}.  For $\beta>\beta^*_0$, the magnetisation $m^{\ast}$ exhibits a discontinuity as $B$ is increased through zero, which indicates that the system is ferromagnetic.  

When the disorder $\eta$ is increased from zero, the critical temperature is reduced (so $\beta^*$ increases).  At a critical disorder strength $\eta_{\rm eq}^{\infty}$,
the critical temperature reaches zero.  (The computation of this critical disorder is given just below.)  For $\eta>\eta_{\rm eq}^{\infty}$, there is no critical point and no ferromagnetic state.

To characterise the behaviour at very low temperatures, we follow \cite{aharony1978tricritical}.
As $\beta \to \infty$, the free energy reduces to the energy, and the minimization condition \eqref{eq:37} becomes
\begin{equation}
  \label{eq:43}
  \mm(h) = 
    \cases{1 & \hbox{for   }  $2m + B + \eta h > 0$ \\
    -1 & \hbox{for   }  $2m + B + \eta h < 0$}
  \; ,
\end{equation}
so (\ref{eq:40}) becomes
\begin{equation}
  \label{eq:44}
  m^{\ast} = 2\int_{0}^{\frac{2m^{\ast}+B}{\eta}} \!\!\! \pdis(h) \, \mathrm{d}h \, .
\end{equation}
For Gaussian disorder then this yields $m^{\ast} = {\rm erf}\left(\frac{2m^{\ast}+B}{\sqrt{2}\eta}\right)$.
To check for ferromagnetism we set $B=0$, in which case ferromagnetic solutions $m^{\ast}\neq0$ exist for $\eta<\eta_{\rm eq}^{\infty}$,
where the critical disorder strength is 
\begin{equation}
  \eta_{\rm eq}^{\infty} = 4/\sqrt{2\pi} \approx 1.596 \; .
\end{equation}
The response to $B$ for $\eta>\eta_{\rm eq}^{\infty}$ is smooth, as one can check by computing $\dst m/\dst h$ from \eqref{eq:44}, see \cite[Eq. (11.35)]{krapivsky2010kinetic}.

Summarizing, the mean-field RFIM with Gaussian disorder exhibits two types of behavior: For weak disorder $\eta<\eta_{\rm eq}^{\infty}$ it resembles the mean-field Ising model, with an associated second-order transition to a ferromagnet.  For strong disorder $\eta>\eta_{\rm eq}^{\infty}$ the ferromagnetic state is destroyed by the disorder and the system is always paramagnetic.

\begin{figure}
  \centering
  \includegraphics[width=0.9\linewidth]{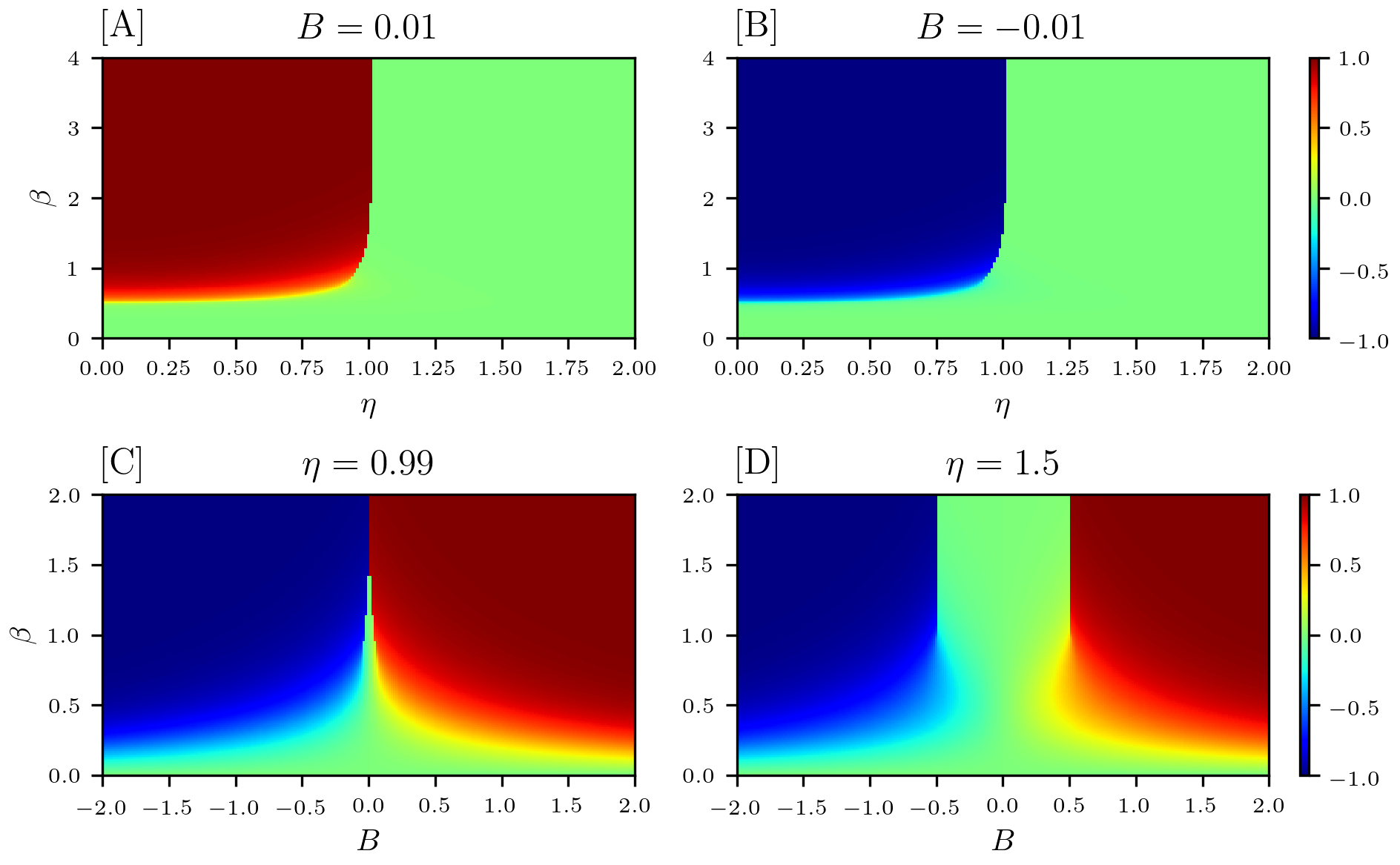}
  \caption{\textbf{Equilibrium phase diagram of the mean-field RFIM with bimodal disorder.} 
   The colors indicate the global magnetisation $m^{\ast}$ (averaged over the disorder), similar to Fig.~\ref{fig:1}.
 \textbf{Panels A and B} show the $(\eta, \beta)$ plane.  
Increasing $\beta$ for $\eta < \eta_{\rm eq}^{(t)}\approx 0.878$ leads to a second-order phase transition where $m^{\ast}$ is continuous.  For $\eta_{\rm eq}^{(t)} < \eta < \eta_{\rm eq}^{\infty}=1$, the magnetisation is discontinuous (as a function of $\beta$), corresponding to a first-order transition. 
\textbf{Panels C and D} show sections of the phase diagram in the $(B,\beta)$ plane for $\eta=0.99$ and $\eta=1.5$ respectively.  In [C], the ferromagnetic transition is first-order.  In [D] there are no singularities at $B=0$ but there are discontinuous transitions when $\beta$ is sufficiently large, which occur at $|B|=B_{\rm eq}^{\ast}=\eta - 1$.
}
  \label{fig:2}
\end{figure}

\subsubsection{Discrete bimodal disorder.}
\label{sec:eqm-bimodal}

The distribution of the random field has a significant impact on the equilibrium phase behavior.  Fig.~\ref{fig:2} shows the phase diagram for discrete bimodal disorder, which may be compared with Fig.~\ref{fig:1}.
We first summarize the similarities between the two cases:
a ferromagnetic phase (where Figs.~\ref{fig:2}[A,B] differ significantly) exists for sufficiently large $\beta$ and sufficiently small $\eta$, similar to Fig.~\ref{fig:1}.  There is a critical value of the disorder $\eta_{\rm eq}^{\infty}$
(see below), and the ferromagnetic transition is lost for $\eta>\eta_{\rm eq}^{\infty} $, in the sense that the magnetisation is continuous at $B=0$, for all $\beta$.

However, there are also significant differences between bimodal and Gaussian disorder.  First, there is a range of $\eta$ in which the transition from paramagnet to ferromagnet is discontinuous (first-order), which is apparent for large $\beta$ in Fig.~\ref{fig:2}[A,B].   Also, Fig.~\ref{fig:2}[D] reveals discontinuities in the magnetisation that appear for $B\neq0$.  A third difference from the Gaussian case (which is not apparent from Fig.~\ref{fig:2} but does affect the underlying computations) is that the system with bimodal disorder supports metastable phases, where $\bar f[{\cal M}]$ has several local minima (which are unrelated by symmetry).

To locate precisely the continuous transition within the small disorder range, one considers the function minimised in Eq.~\eqref{eq:39}: the critical point occurs when the second derivative of this function vanishes at the stationary point $m^{\ast}$. This yields an equation for the critical line (second-order transition line) that reads
\begin{equation}
  \label{eq:41}
  \eta_{\rm eq}(\beta) = \beta^{-1}\arctanh{\sqrt{1 - \frac{1}{2\beta}}} \qquad \hbox{for} \quad \frac{1}{2} \leqslant \beta \leqslant \frac{3}{4} \, .
\end{equation}
The point $\beta_{\rm eq}^{(t)} = 3/4$, $\eta_{\rm eq}(\beta_{\rm eq}^{(t)}) \equiv \eta_{\rm eq}^{(t)} \approx 0.878$ is a tricritical point that separates the line of second order phase transitions ($\beta < \beta_{\rm eq}^{(t)}$) from the first order transition line ($\beta > \beta_{\rm eq}^{(t)}$) \cite{aharony1978tricritical}.


We next consider the situation for very low temperatures $\beta\to\infty$.  Eqs.~(\ref{eq:43},\ref{eq:44}) are still applicable, but there are several possible solutions to Eq.~\eqref{eq:44}, which exist in different (overlapping) ranges of $B$:
\begin{equation}
  \label{eq:45}
  m =
  \cases{
    1 &  \hbox{for  }  $B \geqslant \eta - 2$ \\
    0  & \hbox{for  }  $-\eta \leqslant B \leqslant \eta$ \\
    -1 & \hbox{for  } $B \leqslant -(\eta - 2) $
  }
  \, .
\end{equation}
If multiple solutions exist, the system supports metastable states.  To understand their consequences, fix the magnetic field at some $B\geqslant0$ (and assume $\eta>0$). As long as $\eta - 2 \leqslant B \leqslant \eta$, this means that both $m=1$ and $m=0$ are local minima of the energy.  Comparing the energy of these states, the global minimum is $m=1$ for $0 \leqslant \eta < B+1$ and $m=0$ for $B+1 \leqslant \eta$. 
To obtain the implications for symmetry-breaking transitions, take $B\to0$.   Then one has $m=1$ for $\eta>1$ and $m=0$ for $\eta<1$.  Hence the zero-temperature phase transition to the ferromagnet is of first order, and takes place at
\begin{equation}
  \label{eq:19}
  \eta_{\rm eq}^{\infty} = 1 \, .
\end{equation}
For stronger disorder $\eta > \eta_{\rm eq}^{\infty}$, there is no transition at $B=0$, but states with $m^{\ast}=1$ still exist as local minima of the (free) energy, which can be stabilised by increasing the field $B$.  This leads to a first-order transition at a non-zero field
$B=B_{\rm eq}^{\ast}=\eta - 1$ (see Fig.~\ref{fig:2}[D]), which separates the disordered phase from the magnetised one.

To summarise this section: 
the RFIM with bimodal disorder case displays the same phases as the case of Gaussian disorder, but its phase diagram is richer.  It includes first-order phase transitions that are linked to the existence of metastable states.  
It is quite natural that these metastable states will affect the dynamical large deviations: this will be shown next, in Sec.~\ref{sec:s_ens_RFIM}.

\section{Activity-biased mean-field RFIM}
\label{sec:s_ens_RFIM}

\subsection{Variational form of dynamical free energy}

We now turn to the activity-biased ensemble~\eqref{eq:biased_ensemble} of the mean-field RFIM. 
We adopt the Markov jump dynamics of Sec. \ref{sec:theory:dyn}, where each jump involves a single spin changing its state.
For the mean-field RFIM \eqref{eq:1a}, 
the energy difference for flipping spin $i$ ($\sigma_{i} \to -\sigma_{i}$) is
\begin{equation}
  \label{eq:9}
    E(\bm{\sigma}'|\bm{h}) - E(\bm{\sigma}|\bm{h})   = 2\sigma_{i} \gamma(m,h_i)  - \frac{4}{N} \, \, 
  \end{equation}
where $\bm{\sigma}'$ is the configuration obtained by flipping spin $i$ in $\bm{\sigma}$
and 
\begin{equation}
\gamma(m,h_i) = 2m + B + \eta h_i
\end{equation}
is the local magnetic field.  Hence, the rate for flipping spin $i$ is
\begin{equation}
\label{equ:w1}
\fl\qquad
w_1(\sigma_i,m,h_i) = \chi\left( \beta\gamma(m,h_i) -\frac{2\beta\sigma_i}{N} \right) \exp\left( -\sigma_i \beta \gamma(m,h_i) + \frac{2\beta }{N} \right)
\end{equation}
[Recall that $\chi$ is even in general, and for Glauber dynamics then $\chi(x) = 1/\cosh(x)$.]
The escape rate is $r(\bm{\sigma}) = \sum_i w_1(\sigma_i,m,h_i)$.  

We now construct an ansatz for $\pi_s$ in (\ref{eq:varppl}), based on the mean-field structure of the model.  The guiding principle is that the spins are independent, and their average magnetisation depends on their local field as $\mm(h_i)$, which is a variational function to be optimised.  That is,
\begin{equation}
\label{equ:pi_s_mf}
\pi_s(\bm{\sigma}) = \prod_i \left[ \frac{1+\mm(h_i)}{2}(1+\sigma_i) + \frac{1-\mm(h_i)}{2}(1-\sigma_i) \right]
\end{equation}
For large $N$, this ansatz captures the behavior of the biased ensemble.  In particular using this ansatz in (\ref{eq:varppl}) is sufficient to obtain the exact free energy $\psi$, if $\mm$ is chosen appropriately.
That computation is given in \ref{app:dyn_free_energy}.  The result is that 
\begin{equation}
  \label{eq:21}
  {\psi}(s)  = -{\min}_{\mm} \, \bar{\phi}[\mm, s]
\end{equation}
with
\begin{eqnarray}
  \label{eq:22}
  \bar{\phi}[\mm,s] 
                      & = \bar{r}[\mm] - 2e^{-s}\bar{a}[\mm] \, , 
\end{eqnarray}
where $\bar{r}[\mm]$ is the average escape rate (per spin), and $\bar{a}[\mm]$ is an average mobility.
These are given in turn by
\begin{eqnarray}
  \label{eq:46}
     \bar{r}[\mm] & = \int \pdis(h) r(\mm(h), h, m) \dst h \nonumber \\
     \bar{a}[\mm] & = \int \pdis(h) a(\mm(h), h, m) \dst h
\end{eqnarray}
where $m$ is obtained from $\mm$ using (\ref{eq:m-mm}), and $r,a$ (without overbar) are the local escape rate 
\begin{equation}
  \label{eq:14}
  \fl\qquad
  r(\mm (h),h, m) 
                  =  \chi(\beta \gamma(m,h)) \left[ \cosh(\beta \gamma(m,h)) - \mm(h)\sinh(\beta \gamma(m,h)) \right] \, 
\end{equation}
and mobility
\begin{equation}
  \label{eq:13}
  a(\mm (h),h, m) 
                    =  \frac{1}{2}\chi(\beta \gamma(m,h)) \sqrt{1-\mm(h)^{2}} \, . \\
\end{equation}

To connect the dynamical properties of the model with its equilibrium behavior, it is useful to note that
\begin{equation}
  \label{eq:15}
  r(\mm(h), h, m) = 2 a(\mm(h), h, m) \cosh\left( \frac{\beta}{\pdis(h)} \frac{\delta \bar{f}}{\delta \mm(h)}\right) \, ,
\end{equation}
where $\bar{f}$ is the equilibrium Landau free energy~\eqref{eq:49}.
For $s=0$, the biased ensemble coincides with the equilibrium dynamics and one has from (\ref{eq:partition_function_biased_ens},\ref{eq:def:Psi}) that the dynamical free energy is zero.  To see the connection with (\ref{eq:15}), observe that  for any extremum of $\bar{f}[\mm]$ one has $r(\mm(h), h, m) = 2 a(\mm(h), h, m)$ which indeed yields $\bar\phi[\mm,0]=0$.  It follows that metastable states (and saddles) of the thermodynamic free energy $\bar f$ are all degenerate for the dynamical free energy.

For $s\neq0$, the intuitive meaning of (\ref{eq:22}) is that for positive $s$, the system tends to minimise the escape rate $\bar r$ (hence reducing the activity) while for negative $s$, it tends to enhance the mobility $\bar a$ (which increases the activity).

For the discrete bimodal disorder distribution, we recall that the function $\mm$ is completely specified by two numbers which are the magnetisations of the subsets of spins with $h_i=\pm1$.  Hence the minimization in (\ref{eq:21}) can be performed numerically, to obtain the dynamical free energy.  That calculation will be performed in Sec.~\ref{sec:full_phase_diag_activity}.  Before that, we address some asymptotic regimes, using arguments that are relevant for generic $\pdis$.

\subsection{Asymptotic regimes of the activity-biased mean-field RFIM}
\label{sec:asymp-dyn}

We discuss the behavior of $\psi$ and $\bar\phi$ in three asymptotic regimes: the limits of low- and high-activity, and the behavior for small bias $|s|\ll 1$.
This analysis will identify the main physical principles which control the behavior of biased ensembles.

\subsubsection{Low activity regime $s\gg 1$.}\label{sec:low_act_reg}

For biased ensembles with very low activity $s\to\infty$, minimisation of \eqref{eq:22} reduces to minimisation of the escape rate $\bar{r}$.  (Since there are very few spin flips, trajectories remain for long periods in single configurations, and the least unlikely mechanism for this is to find configurations where $r(\bm{\sigma})$ is small.)

For the Glauber rule with a microscopic mobility $\chi(x) = 1/\cosh(x)$, the dynamical Landau free energy from Eq.~\eqref{eq:14} is
\begin{equation}
  \label{eq:47}
  \lim_{s\to +\infty} \bar{\phi}[\mm, s] = 1 - \int \pdis(h) \mm(h) \tanh(\beta \gamma(m,h)) \dst h\, .
\end{equation}
This is a functional of ${\cal M}$ whose local minima obey~\eqref{eq:43}, which was derived above by local minimisation of the energy.  
In other words, local energy minima are also local minima of the escape rate (and vice versa).
This demonstrates a link between the behaviour of the system in two separate limits: $s\to \infty$ (with fixed $\beta$) and $\beta\to \infty$ (with fixed $s=0$).

For Gaussian disorder, the large-$s$ behavior can then be deduced from Fig.~\ref{fig:1}: 
for $\eta \leqslant \eta^\infty_{\rm eq}$ [as given by (\ref{eq:44})] then ferromagnetic states are dominant as $s\to\infty$.  For stronger $\eta \geqslant \eta^\infty_{\rm eq}$ the large-$s$ behaviour for $B=0$ has $m=0$, but there is a transition to finite magnetisation at sufficiently large $B$.

\begin{figure}
  \centering
  \includegraphics[width=\linewidth]{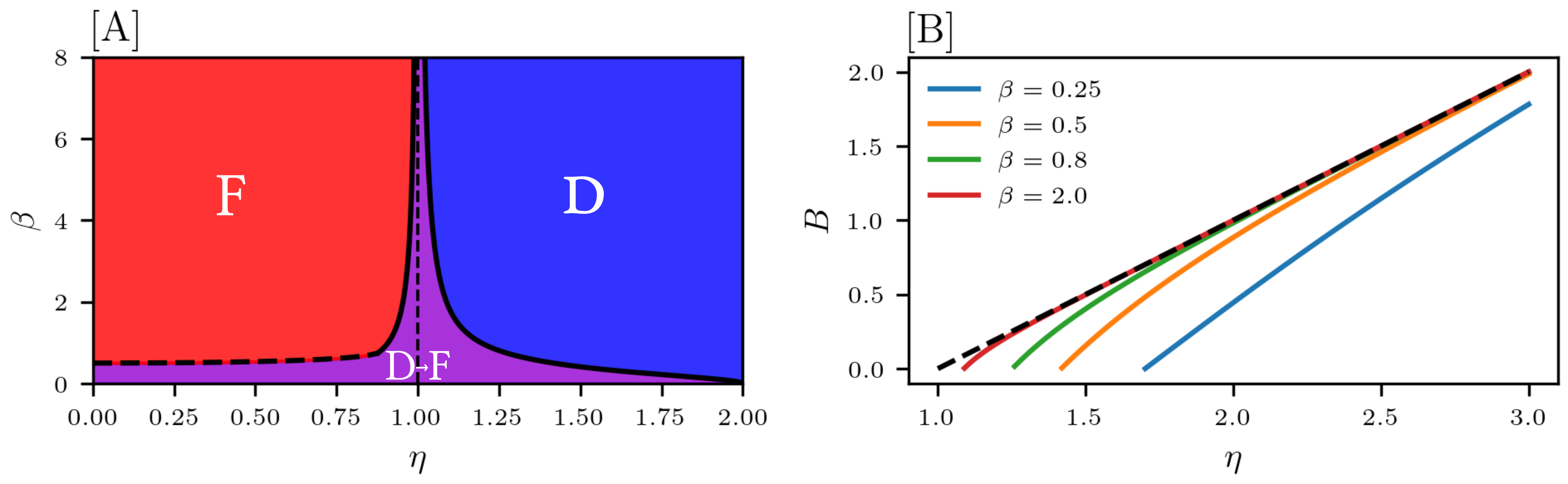}
  \caption{\textbf{The low activity limit ($s\to \infty$) for the RFIM with discrete bimodal disorder.} 
  \textbf{[A]}~For $B=0$, we compare the equilibrium phases with the states of minimal escape rate.  Region F (red) indicates the equilibrium ferromagnetic phase, as in Fig.~\ref{fig:2} (in terms of biased ensembles, this region means that $m\neq0$ at $s=0$.  Region D (blue) indicates the range of parameters for which the minimal-activity state is disordered [$\bar{r}_{\rm D}<\bar{r}_{\rm F}$ in (\ref{eq:16})], so $m=0$ as $s\to\infty$).  In the D-F region (purple), the system is disordered for $s=0$ but ferromagnetic for $s\to\infty$, showing that a phase transition must take place for some finite $s$.  
\textbf{[B]}~Colored lines show the external magnetic field $B^{\ast}(\eta)$ such that $\bar{r}_{\rm F}=\bar{r}_{\rm D}$.  Above these lines (which depend on the temperature), one has ferromagnetic behaviour as $s\to\infty$.  Above the dashed black line, the global energy minimum is ferromagnetic (independent of temperature).  Between the dashed and colored lines, the global minimum of the escape rate is ferromagnetic while the global energy minimum is disordered.  At low temperature $\beta\to\infty$, the minima coincide.}
  \label{fig:3}
\end{figure}

For bimodal disorder, the situation is slightly different.  Local minima of the energy are also minima of the escape rate $\bar r$, but this does not guarantee that the global minimum of $\bar r$ is the global energy minimum.  The relevant local minima are given by (\ref{eq:45}).  Their escape rates are
\begin{eqnarray}
  \label{eq:16}
    \bar{r}_{\rm F} &= 1 - \frac{1}{2}\Big[\tanh\left[\beta(2+B+\eta)\right] + \tanh\left[\beta(2+B-\eta)\right] \Big] \\
        \bar{r}_{\rm D} &= 1 - \frac{1}{2}\Big[\tanh\left[\beta(B+\eta)\right] - \tanh\left[\beta(B-\eta)\right] \Big] \, .
\end{eqnarray}
which correspond to $m=1$ (ferromagnetic) and $m=0$ (disordered) states, respectively.
The resulting behavior at $B=0$ is summarized in Fig.~\ref{fig:3}[A], which indicates which state has minimal escape rate, and how this compares with the equilibrium state. For strong disorder $\eta>1$, the minimal-activity state is disordered for $B=0$, but increasing the field leads to a discontinuous transition to a magnetized state at some $B=B^{\ast}(\eta)$.
Fig.~\ref{fig:3}[B], shows the behavior of $B^{\ast}(\eta)$, which illustrates clearly that the global minimum of the energy does not always coincide with that of the escape rate.

\subsubsection{High activity regime $s\ll -1$.}\label{sec:high_act}
For large negative bias (high activity) one sees from~\eqref{eq:22} that the system acts to maximize the mobility $\bar a$ 
which can be expressed for Glauber dynamics [using \eqref{eq:13}] as
\begin{equation}
  \label{eq:28}
  \bar{a}[\mm] = \frac{1}{2} \int \pdis(h) \frac{\sqrt{1-\mm(h)^{2}}}{\cosh(\beta \gamma(m,h) )} \dst h \, .
\end{equation}
The numerator is an entropic term that is maximal when $\mm(h)=0$ (so spins are equally likely in either state) and the denominator favors states where the local fields $\gamma(m,h)$ are small in magnitude.

For the special case $B=0$, the disordered state where $\mm(h)= 0$ for all $h$ is always an extremum of $\bar{a}$.
This achieves the maximal possible value for the numerator in \eqref{eq:28}, and for small $\eta$ the effect of the denominator is small, so one expects this disordered state to maximize $\bar a$ quite generally, as in the pure case (without disorder)~\cite{GuiothJack2020}.  However, for larger $\eta$, the denominator becomes important and the maximum of $\bar a$ requires minimization the local fields $\gamma$.  For discrete bimodal disorder, \ref{app:max_mob} shows that this leads to ferromagnetic behaviour, when $\eta$ is sufficiently large.


For $B\neq0$, an interesting result from the pure case~\cite{van2010second, GuiothJack2020} is that maximal mobility is achieved when the magnetisation of the system is opposite to the applied field.  We will show below that the same situation arises in the RFIM (when $\eta$ is not too large).  The reason is that taking $m$ antiparallel to $B$ helps minimise the local field $\gamma$.

\subsubsection{Close to equilibrium regime $|s|\ll 1$}\label{sec:close_to_eq_reg}

As already encountered for the pure Ising model~\cite{GuiothJack2020}, the typical states of the activity-biased system for $|s|\ll1$ are closely related to the structure of the equilibrium Landau free energy.  
At first order in $s$, the dynamical Landau free energy~\eqref{eq:22} reads
\begin{equation}
  \label{eq:18}
  \bar{\phi}[\mm, s] = \bar{\phi}_{\rm eq}[\mm] + 2s\bar{a}[\mm] + \mathcal{O}(s^{2}) \,
\end{equation}
with $\bar{\phi}_{\rm eq}[\mm] = \bar{\phi}[\mm, 0]$.
Recalling (\ref{eq:15}) and the associated discussion, $\bar{\phi}_{\rm eq}[\mm]=0$ whenever $\mm$ extremises the equilibrium free energy $\bar f$,  and $\bar{\phi}_{\rm eq}[\mm] >0$ otherwise.  
This means that the $\bar\phi_{\rm eq}[\mm]$ has degenerate minima whenever $\bar f$ is not convex.

In such cases, the dominant effect of small $s$ is to break the degeneracy: the minimum of $\bar\phi$ is achieved by the extremum of $\bar f$ with the minimal value of $s\bar a$.
This offers a mechanism for dynamical phase transitions, which can be related to equilibrium properties.  

For the RFIM with weak disorder, one expects a situation similar to the pure case.  The free energy is convex at  high temperatures, so the response to the biasing field $s$ is smooth near $s=0$.  However, for temperatures below criticality (and sufficiently small $B$), the free energy has two minima (corresponding to ferromagnetic states), separated by a saddle point.  In all cases analysed here, the ferromagnetic states have lower mobility $\bar a$ so the response for small positive $s$ is smooth, with the system remaining ferromagnetic.  On the other hand, for small negative $s$, the dynamical free energy $\bar\phi$ is minimised by a state that is localised near the saddle point of the equilibrium free energy.  For $B=0$ the saddle has $m^{\ast}=0$ so one has a discontinuity in $m^{\ast}$ between $s\geq0$ (ferromagnetic minimum) and $s<0$ (saddle point).  This corresponds to a first-order dynamical phase transition.  Physically, the origin of this transition is that there is no thermodynamic driving force that pushes the system away from the saddle, so systems may remain localized there with (relatively) high probability.  This is an effective mechanism for increased activity.

For stronger disorder, we recall from Sec.~\ref{sec:eq-phase} that the equilibrium free energy surface can become more complicated. In particular, for discrete bimodal disorder and $\eta_{\rm eq}^{(t)} < \eta < \eta_{\rm eq}^{\infty}$, the equilibrium free energy has three minima, corresponding to two ferromagnetic states as well as a paramagnetic one.  
The ferromagnetic states have the smallest mobility so they minimize $\bar\phi$ for small positive $s$.  For small negative $s$, the mobility $\bar a$ is maximized (and hence $\bar\phi$ is minimized) by one of the saddle points of the equilibrium free energy.  For the specific case $s=0$, the behavior is determined by minimum of the equilibrium free energy, which may be a ferromagnetic state, or a paramagnet, depending on $\eta$.  Overall, the result is a first-order phase transition at $s=0$, with the phases for positive and negative $s$ determined by their mobilities, through (\ref{eq:18}).

\subsection{Full $(s,B)$-phase diagrams for bimodal disorder}\label{sec:full_phase_diag_activity}

We now turn to the dynamical phase diagrams for the RFIM in the case of discrete bimodal disorder.
In this case ${\cal M}(h) = m_+ \delta(h-1) + m_- \delta(h+1)$ so the functional minimization of (\ref{eq:21}) reduces to a simple minimization over $(m_+,m_-)$ which is performed numerically.
%
%
Our primary focus is on ensembles with lower-than-average activity ($s>0$), since this is the regime that is most relevant for the analogy with glass-forming systems, as discussed in Sec.~\ref{sec:glass}.  


We recall from Fig.~\ref{fig:2} and the accompanying discussion that the equilibrium phase behavior of this system has three regimes, according to the strength of the disorder.  Since the behavior of biased ensembles for $s\approx 0$ is controlled by the equilibrium phases, we separate these three regimes.

\begin{figure}
  \centering
  \includegraphics[width=0.9\linewidth]{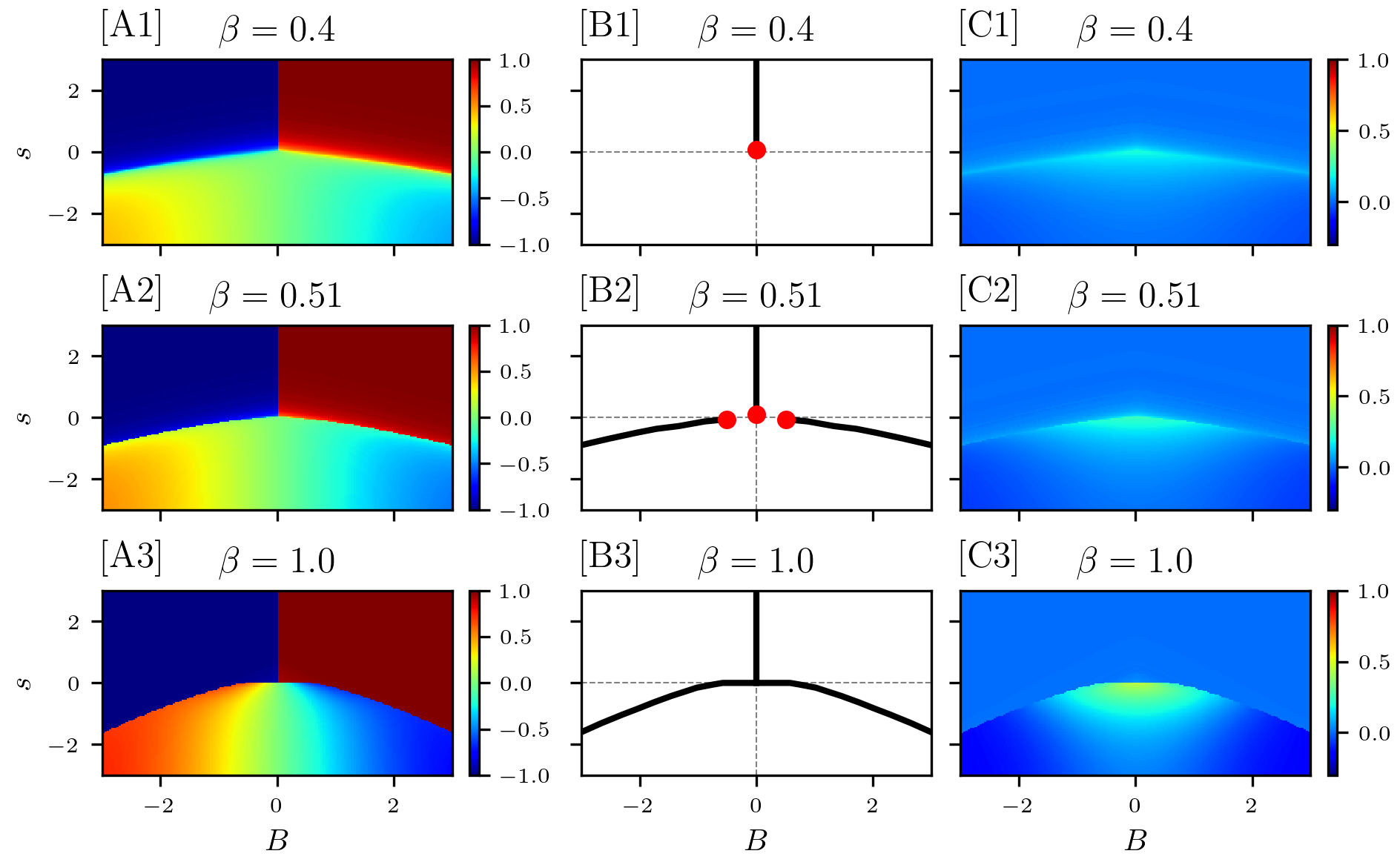}
  \caption{\textbf{Dynamical phase behaviour (weak disorder)} for the RFIM with discrete bimodal disorder. The disorder strength is fixed at $\eta=0.5$. 
  \textbf{[A]} Average magnetisation $m$ in the biased ensembles. 
  \textbf{[B]} Locations of first-order transition lines (black) and critical points (red dots), as estimated from [A]. 
  \textbf{[C]} Average overlap $\tau$ in the biased ensemble.}
  \label{fig:4}
\end{figure}

The dynamical phase behaviour depends on four parameters $(\eta,\beta,B,s)$.  Each regime is illustrated by a representative value of $\eta$; we then select illustrative values of $\beta$ and plot results in the $(B,s)$-plane.
Specifically, we plot the average magnetization $m$ and overlap $\tau$ in the biased ensemble, as functions of $(B,s)$.  We also show diagrams that illustrate where singularities occur.  As a reference point, it is useful to recall the behaviour shown in Figs.~\ref{fig:1},\ref{fig:2} as a function of $(B,\beta)$: we already noted above that the behaviour for large positive $s>0$ (minimal escape rate) shares some features with large $\beta$ (minimal energy).

\subsubsection{Weak disorder.}\label{sec:small_dis}

The regime of weak disorder is $\eta \leqslant \eta_{\rm eq}^{(t)}$.  Phase diagrams are displayed in Fig.~\ref{fig:4} for the representative case $\eta=0.5$.   Recalling Fig.~\ref{fig:2}, the equilibrium phase behaviour has a critical point at $\beta_{\rm eq}^{\ast}(\eta = 0.5) \approx 0.537$.  We show results for  the high-temperature regime ($\beta=0.4$); a paramagnetic regime that is close to the critical point ($\beta=0.51$); and the low temperature ferromagnetic regime.

The qualitative results are the same as for the pure Ising model discussed in~\cite{GuiothJack2020}, to which we refer for a detailed discussion.  We explain them here using the arguments of of Sec.~\ref{sec:asymp-dyn}.  For large positive $s$, one finds states of minimal escape rate, which are ferromagnetic.  For large negative $s$, one finds states of maximal mobility, which have their magnetisation oriented opposite to the applied field $B$ (in order to minimize the local field).  Close to $s=0$, the situation depends on the equilibrium free energy: For high temperatures $\beta<\beta_{\rm eq}^{\ast}$, the equilibrium free energy is convex so the magnetization is analytic at $s=0$, although there are critical points and phase transitions elsewhere in the phase diagram~\cite{GuiothJack2020}.  At low temperatures then the equilibrium free energy has two ferromagnetic minima and a saddle point: hence there are three different dynamical phases, which are localised near these extrema of $\bar f$.  The phases coexist at $(B,s)=(0,0)$, with ferromagnetic phases appearing for $s>0$ and the saddle-point being selected for $s<0$.
{Finally, the presence of (horizontal) first order transition lines for $s<0$ and  $B\neq 0$ appears as a non-trivial output of the interplay between the escape rate and the mobility of the trajectories. More details are available in \cite{GuiothJack2020}.}

Based on these arguments, we expect that this general picture also holds for the case of Gaussian disorder.

\subsubsection{Intermediate disorder.}
\label{sec:inter_dis_activity}

\begin{figure}
  \centering
  \includegraphics[width=0.9\linewidth]{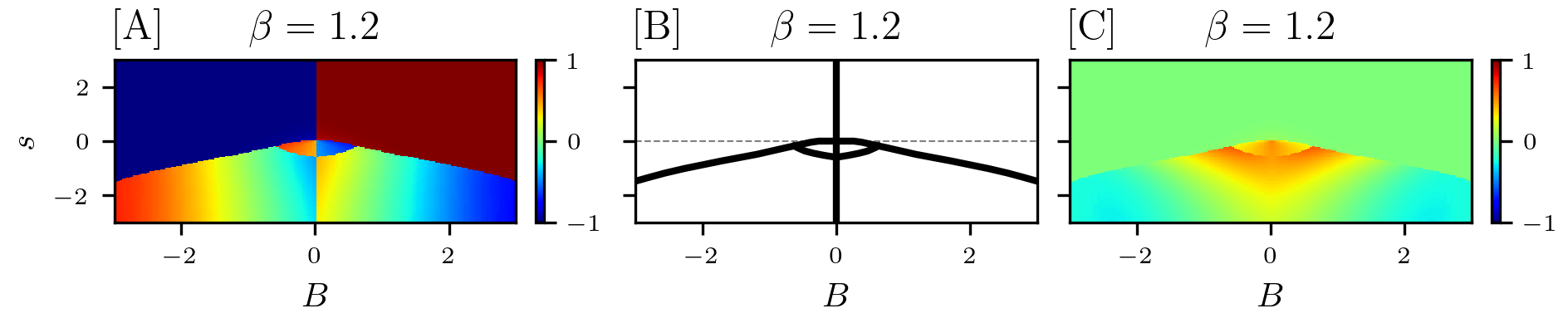}
  \caption{
  \textbf{Dynamical phase behaviour (intermediate disorder)} for the RFIM with discrete bimodal disorder. The disorder strength is fixed at $\eta=0.95$. \textbf{[A]} Average magnetisation $m$ in the biased ensemble. 
  \textbf{[B]} Locations of first-order transition lines (black), as estimated from [A]. 
  \textbf{[C]} Average overlap $\tau$ in the biased ensemble.}
  \label{fig:5}
\end{figure}

As discussed in Sec.~\ref{sec:eqm-bimodal}, the RFIM with bimodal disorder supports metastable states in the 
regime of intermediate disorder $\eta_{\rm eq}^{(t)} \leqslant \eta \leqslant \eta_{\rm eq}^{\infty}$ (and recall $\eta_{\rm eq}^{\infty}=1$).  This has consequences for dynamical phase behavior, particularly for weak biasing.  
Fig.~\ref{fig:5} shows the behavior for $\eta=0.95$ at a temperature $\beta=1.2$ which is in the ferromagnetic regime [$\beta>\beta_{\rm eq}(\eta)$], and the paramagnetic metastable state still exists.  

Discussing first the asymptotic regimes: for $s\gg 1$ (minimal escape rate) the system is ferromagnetic.  For $s\ll -1$ (maximal mobility), the behaviour is again ferromagnetic, as discussed in Sec.~\ref{sec:high_act} and \ref{app:max_mob} (this differs from the behavior at weak disorder).  For small $s$, we examine the extrema of the free energy, which are the two ferromagnetic states, the paramagnetic metastable state, and (at least) two saddle points.  The ferromagnetic states are the extrema of $\bar{f}$ with minimal mobility, so they are selected when $s$ is small and positive.  For small negative $s$, it turns out that the mobility is maximised at the saddle points, which have finite magnetisation. The dominant saddle has its magnetisation opposite to $B$, which reduces the local field and enhances the mobility.  As a result, the point $(B,s)=(0,0)$ involves coexistence of four different dynamical phases, corresponding to two minima and two saddles of $\bar f$.  
[Such ``quadruple points'' are not expected in equilibrium phase diagrams, but they can occur in this dynamical context because all extrema of $\bar f$ have $\bar\phi[\mm,0]=0$, and may correspond to coexisting phases.]

The detailed structure of the phases for $s<0$ is quite complex, due to the interplay between metastable states and saddles of the free energy surface.  The general expectation -- which is also true here -- is that for small negative $s$ then the behaviour is controlled by the  maximally active extrema of the (static) free energy, which are typically saddle points; for large negative $s$ the system acts to maximize its activity, independent of the free energy.

\subsubsection{Strong disorder.}
\label{sec:strong_dis_activity}

\begin{figure}
  \centering
  \includegraphics[width=0.9\linewidth]{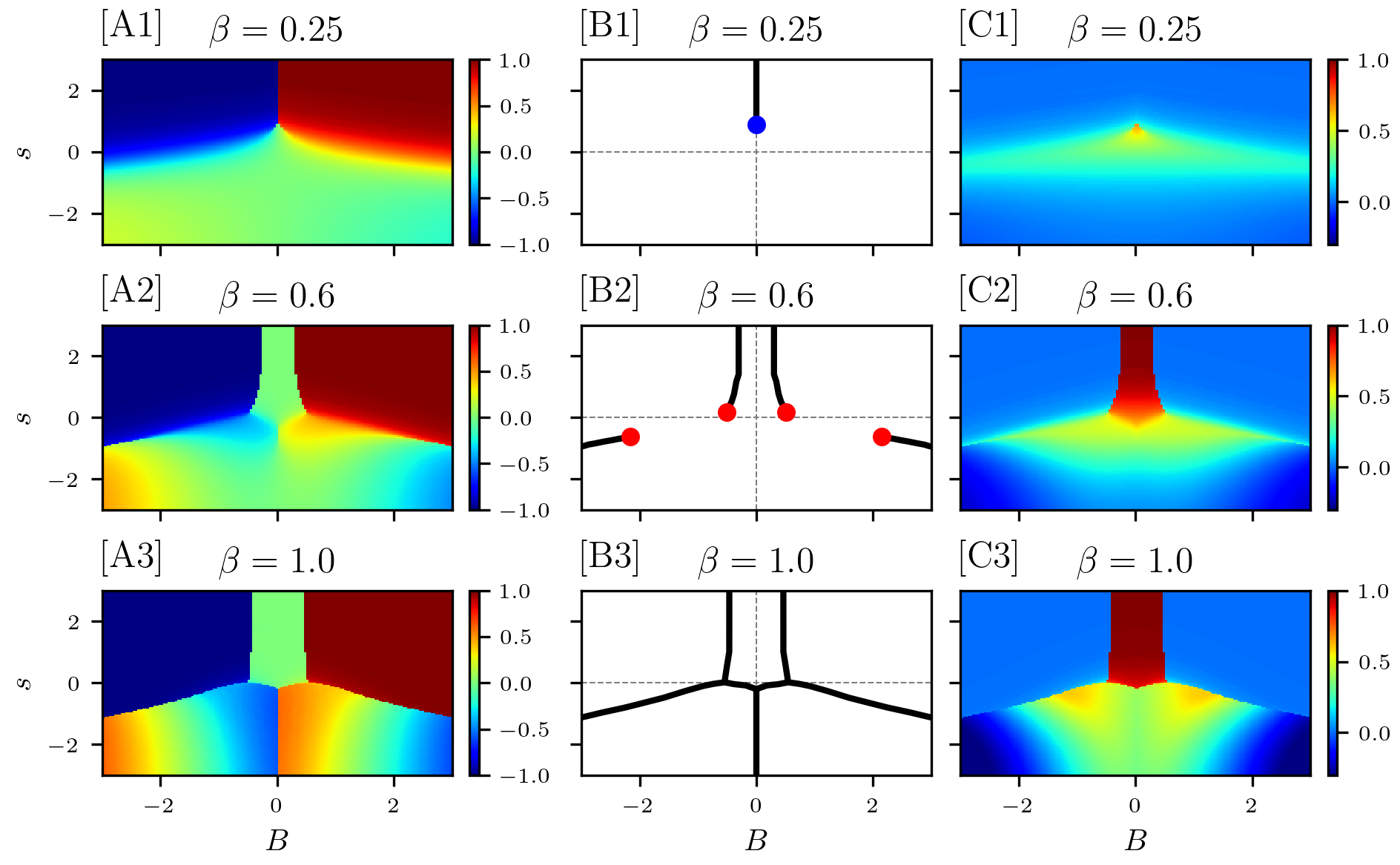}
    \caption{\textbf{Dynamical phase behaviour (strong disorder)} for the RFIM with discrete bimodal disorder. The disorder strength is fixed at $\eta=1.5$. 
  \textbf{[A]} Average magnetisation $m$ in the biased ensembles. 
  \textbf{[B]} Locations of first-order transition lines (black), critical points (red dots), and triple points (blue dot), as estimated from [A]. 
  \textbf{[C]} Average overlap $\tau$ in the biased ensemble.}
  \label{fig:6}
\end{figure}

Finally, Fig.~\ref{fig:6} shows the behaviour for the strong disorder regime $\eta > \eta_{\rm eq}^{\infty}$, specifically $\eta=1.5$.  To understand these results,  recall from Fig.~\ref{fig:2} that the ferromagnetic phase transition is destroyed by the disorder in this case, but that first-order phase transitions do occur with $B\neq0$, at sufficiently low temperatures.  Also, Fig.~\ref{fig:3} shows which states minimize the escape rate: for $B=0$ these may be  ferromagnetic (at small $\beta$) or disordered (large $\beta$).  These states dominate as $s\to\infty$.

Fig.~\ref{fig:6}[A1] shows a high-temperature case ($\beta=0.25$), showing ferromagnetic behavior at large $s$, as expected.  For $s\ll-1$ the behaviour is disordered.  The free energy $\bar f$ is convex, so the dynamical free energy behaves analytically at $s=0$.  The behaviour for $s>0$ resembles the equilbrium phase diagram shown in Fig.~\ref{fig:2}[C], which has $\eta=0.99$, smaller than the current value $\eta=1.5$.  Hence, the effect of the increasing $s$ in the biased ensemble is qualitatively similar to an equilibrium system where one simultaneously increases $\beta$ and reduces $\eta$.  That is, a bias to low activity tends to favour more ordered states with lower energy, consistent with physical intuition.

Fig.~\ref{fig:6}[A2] shows intermediate temperature ($\beta=0.6$).  In this case the ferromagnetism is destroyed by the disorder and we find $m=0$ at $(B=0,s\to\infty)$.  For $s>0$, the behavior resembles Fig.~\ref{fig:2}[D]: in this case, increasing $s$ in the biased ensemble is similar to increasing $\beta$ at equilbrium.  In other words, a bias to low activity favours states with lower energy.  The behavior for $s<0$ is broadly similar to Fig.~\ref{fig:4}[B2].  

  Fig.~\ref{fig:6}[A3] shows a low temperature case ($\beta=1.0$). The equilibrium system has first-order phase transitions at $B=\pm B^{\ast}$ (recall Fig.~\ref{fig:2}), which appear in the dynamical phase diagram as phase transitions with $s=0$.  These phase transitions extend into the positive half-plane ($s>0$) because the paramagnetic and ferromagnetic (equilibrium) phases have relatively low activity and also minimize the dynamical free energy (the ferromagnetic phase is slightly favored for larger $s$).  For sufficiently large negative $s$ a new dynamical phase transition appears at $B=0$, similar to that observed for intermediate disorder in Fig.~\ref{fig:5}.  

\section{Connection to glassy systems : RFIM with broken spin-flip symmetry}
\label{sec:asym-mob}

\subsection{The different symmetries of RFIM and RFOT systems}

We now discuss the connection between the activity-biased RFIM as studied in Sec.~\ref{sec:s_ens_RFIM} and the effective theory of the glass transition.
The RFOT phase diagram of Fig.~\ref{fig:0} is based on a free energy functional whose order parameter is the overlap $q(\mathcal{C},\mathcal{C}_{\rm ref})$ between two microscopic configurations $\mathcal{C}, \mathcal{C}_{\rm ref}$.  [The overlap is large ($q\approx 1$) when the configurations are similar, and small ($q\approx 0$) for uncorrelated configurations.]
The reference configuration $\mathcal{C}_{\rm ref}$ acts as a source of quenched disorder.  The dominant terms of the corresponding free energy functional are~\cite{franz2013glassy,biroli2014random} 
\begin{equation}
  \label{eq:48}
  \mathcal{V}[q|\mathcal{C}_{\rm ref}] = \int  \big[ c |\nabla q(x)|^{2} + V(q(x)) + h_{\mathcal{C}_{\rm ref}}(x)q(x) - \epsilon q(x)\big]  \dst x\, ,
\end{equation}
where $V$ is an asymmetric double-well potential, $c$ is a stiffness coefficient, and the term with $h_{\mathcal{C}_{\rm ref}}$ encapsulates the coupling of the quenched disorder to the overlap.   If one makes a mean-field approximation (ignoring the spatial structure), this free energy can be considered analogous to the RFIM free energy \eqref{eq:49}, with the correspondences of Table~\ref{tab:translation_table} (from the Introduction).  However, this free energy lacks the simple symmetry of the RFIM case (whose free energy is invariant under $m\to-m$, for $B=0$).

Given these facts, a natural question follows: For a system with free energy \eqref{eq:48}, how is the dynamical phase behaviour different from an RFIM with free energy \eqref{eq:49}?  For a general analysis of dynamical properties, one requires an assumption for the dynamical free energy \eqref{eq:22}, which amounts to an assumption for the mobility.  Detailed predictions for the mobility are not available, but some insight into the expected behavior is available by simple symmetry arguments.

As already shown in Tab.~\ref{tab:translation_table}, the absence of symmetry in the RFOT case means that the line of (equilibrium) phase coexistence becomes a non-trivial function $\epsilon_{\rm eq}(T)$.  Tab.~\ref{tab:translation_table} and Fig.~\ref{fig:0} indicate that part of this effect can be accounted for by identifying the RFIM field $B$ with $\epsilon - \epsilon_{\rm eq}(T)$. However, 
the crucial observation for dynamical behaviour is that if the system lacks spin-flip symmetry, coexisting phases should have different dynamical activities.  Hence, a bias $s$ will drive the RFOT system towards one of the coexisting phases.  
Indeed, the RFOT (glassy) context is one where the high overlap (glass) phase has much lower dynamical activity than the liquid, so that positive bias $(s>0)$ favours the glass, as in~\cite{hedges2009dynamic}. This contrasts with the RFIM at $B=0$, where Figs.~\ref{fig:4} and~\ref{fig:5} show that $s>0$ favours ferromagnetic phases, but without any preference for positive or negative magnetisation.

To gain insight into this effect, we 
analyse a version of the RFIM where the free energy remains the same, but the mobilities of the coexisting phases are different.  The result is that the bias naturally drives the system towards ferromagnetic phases with $m>0$, which we identify with the high-overlap (glass) phase in the RFOT context.  The connections of this result to glassy systems are discussed in Sec.~\ref{sec:discuss}.

\subsection{An RFIM with asymmetric mobility}

We consider an RFIM as above, but now with
asymmetric mobility $\chi_{a}$:
\begin{equation}
  \label{eq:8}
  \chi_{a}(m,h) = (1 - \mu m ) \chi(\beta \gamma(m,h)) \, ,
\end{equation}
with $\mu\in[-1,1]$ a real parameter, so that $\mu \neq 0$ induces asymmetry in the mobility.  We will restrict to the case $\mu\in [0,1]$ so the states with positive $m$ have lower mobility (similar to the RFOT case where one expects lower mobility for $q>0$).
From Eqs~\eqref{eq:13} and~\eqref{eq:22}, the corresponding dynamical free energy $\bar{\phi}_{a}$ is simply
\begin{equation}
  \label{eq:1}
\bar{\phi}_{a}(\mdis, s) = (1-\mu m)\bar{\phi}(\mdis,s) \;.   
\end{equation}

\begin{figure}
  \centering
  \includegraphics[width=0.9\linewidth]{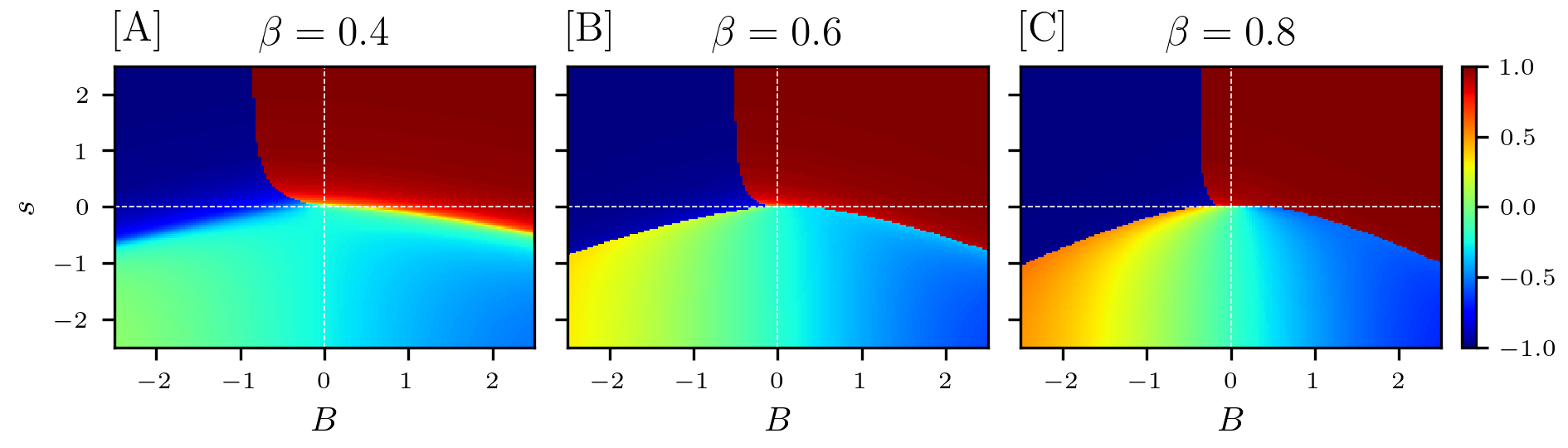} 
  \caption{\textbf{Dynamical phase behaviour (asymmetric mobility, weak disorder)} for the mean-field RFIM with bimodal disorder. The disorder strength is fixed at $\eta=0.5$. Parameters are:  $\mu=0.5$,  $\eta=0.5$, $\beta=0.4$ [A], $\beta=0.6$ [B], $\beta=0.8$ [C].
}
  \label{fig:7}
\end{figure}

\begin{figure}
  \centering
  \includegraphics[width=0.9\linewidth]{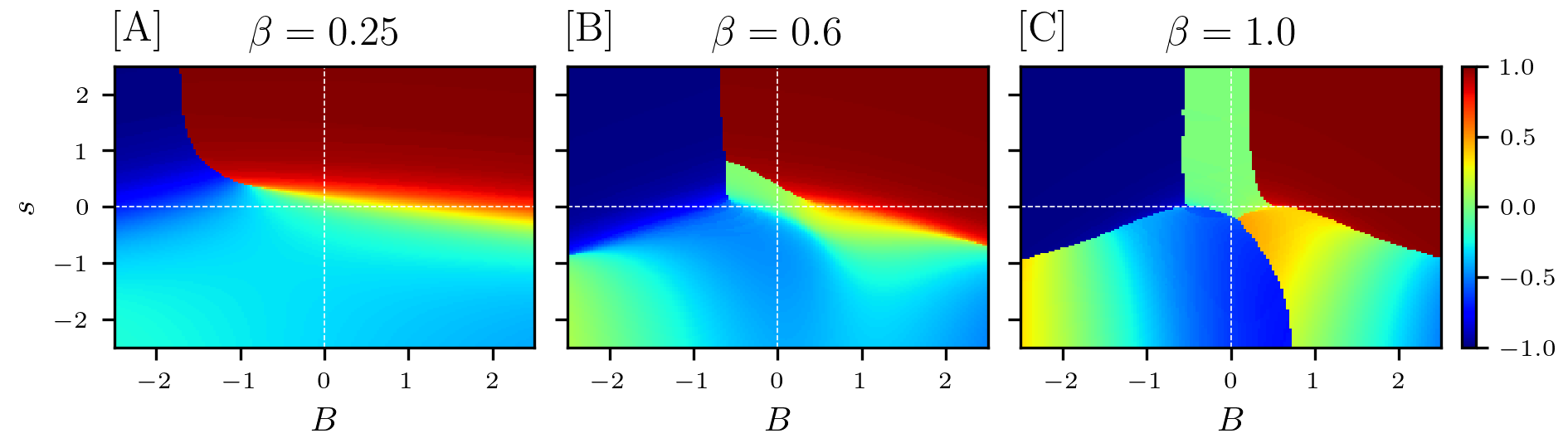} 
  \caption{\textbf{Dynamical phase behaviour (asymmetric mobility, strong disorder)} for the mean-field RFIM with bimodal disorder.  The disorder strength is fixed at $\eta=1.5$. Parameters are:  $\mu=0.5$,  $\eta=1.5$, $\beta=0.25$ [A], $\beta=0.6$ [B], $\beta=1$ [C].
}
  \label{fig:8}
\end{figure}

A similar numerical analysis as the one used for the symmetric mobility case in Sec.~\ref{sec:s_ens_RFIM} can be performed.
As an illustrative case, we consider bimodal disorder and $\mu=0.5$, with results in
Figs.~\ref{fig:7} and \ref{fig:8}.

In particular, 
Fig.~\ref{fig:7} shows the dynamical phase behaviour in the regime with weak disorder ($\eta=0.5<\eta_{\rm eq}^{\infty}$), for $\mu=0.5$. 
These results 
use the same temperatures as in Fig. \ref{fig:4}. 
As expected, the first-order dynamical phase transition for $s>0$ no longer follows the line $B=0$, because the spin-reversal symmetry has been broken.  Instead, positive bias $s>0$ selects positive magnetisation for states on that line.  If one instead takes small $B<0$, the magnetisation increases with $s$.  Depending on the temperature, this may happen as a smooth crossover or a first-order transition, the relevant theory is similar to the symmetric case (Sec.~\ref{sec:small_dis}).

It is notable that this effect is also present in the absence of disorder ($\eta=0$).   For example, consider the large-$s$ limit as in Sec.~\ref{sec:low_act_reg}, so that the free energy is controlled by the escape rate $\bar{r}$.  This function has two local minima, corresponding to the two ferromagnetic states.  Dynamical phase coexistence happens when the two minima have equal values of $\bar{r}$: a positive bias $s>0$ favours positive magnetisation so phase coexistence requires $B<0$ to compensate.  As $s\to\infty$, coexistence occurs at finite $B=B^*$, dependent on $\mu,\beta$.  (At very low temperatures $B^*\to0$, this is because the asymmetry of the mobility is assumed independent of temperature, while the field $B$ enters the dynamics as $\beta B$, so the effect of $B$ dominates the asymmetry $\mu$ at low temperatures.)

Fig.~\ref{fig:8} shows corresponding results for strong disorder $\eta=1.5 > \eta^{\infty}_{\rm eq}$, which are comparable with Fig.~\ref{fig:6}.
As in the case of weak disorder, the phase boundaries no longer follow symmetry lines, but the structure of the phase diagram is similar.
However, the behaviour for $s\to\infty$ is different for some temperatures (for example $\beta=0.6$), in that one finds a phase transition between two ferromagnetic states (recall that Fig.~\ref{fig:6}[B] has two ferromagnetic states separated by a disordered phase, when $s$ is large).  
The reason for this difference is that the combination of $s>0$ and $\mu>0$ favours the positive-magnetisation state over the disordered one.

\subsection{Discussion}
\label{sec:discuss}

\begin{figure}
  \centering
  \includegraphics[width=0.9\linewidth]{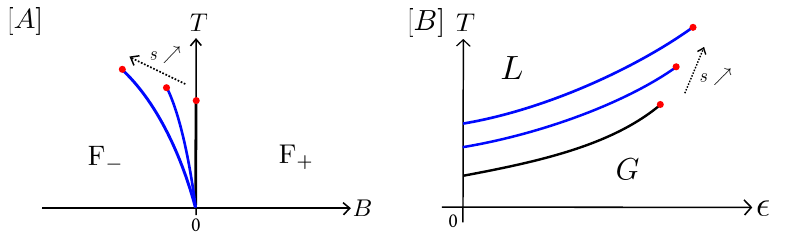}
  \caption{Sketches of the phase diagrams of the (weak disorder) asymmetric mobility RFIM [\textbf{A}] and the replica-biased glass-forming system [\textbf{B}] in the $(T,B)$ or $(T,\epsilon)$ parameter space for different activity-bias $s>0$. [\textbf{A}]: the blue lines indicate first order transitions between the two ferromagnets when increasing $s>0$, as read on the phase diagrams from Fig. \ref{fig:7}. The lines terminate by a second order phase transition for large $T$. [\textbf{B}]: Sketch of the corresponding phase diagram for a glass-forming system in the $(T,\epsilon)$ plane, according to Tab.~\ref{tab:translation_table}.}
  \label{fig:9}
\end{figure}

We discuss how these results for RFIMs with asymmetric mobility would manifest in RFOT systems, based on the analogy of Fig.~\ref{fig:0} and Table~\ref{tab:translation_table}.
We focus on $s>0$, where the behaviour is simpler, and the connection to glassy systems is relevant.
Fig.~\ref{fig:9} shows one representation of the effect of positive bias $s$, in the regime of weak (or moderate) disorder shown in Fig.~\ref{fig:0}.  For the RFIM we consider the case $\mu>0$; in the RFOT case, we assume that models with high overlap will have lower dynamical activity (because such systems remain localised near the reference configuration, instead of exploring the ergodic fluid state).  In contrast to the main text, we show phase diagrams as a function of temperature, instead of the inverse temperature $\beta$.  

Fig.~\ref{fig:9}[A] shows that the effect of the bias $s>0$ is to shift the RFIM critical point to higher temperature, and also to negative field.
Note that a more precise asymptotic estimate can be found by minimising the dynamical Landau free energy~(\ref{eq:1}) (with a similar analysis as in Sec. \ref{sec:high_act}): the coexistence line behaves as $T^{\ast}(B)\sim B/\mu$ in the limit $s\to\infty$, where the critical temperature $T_c\to +\infty$ along this direction.
Physically, this is easily understood because the bias favours ordered states and positive magnetisation, which are both associated with lower dynamical activity.

Fig.~\ref{fig:9}[B] shows the corresponding situation for the RFOT phase diagram (assuming that the disorder is weak enough that the equilibrium phase transition is present).  Broadly speaking, the effect of a positive bias $s>0$ should be to stabilise the glass state (that has the lowest escape rate).  By analogy with the asymmetric activity-biased RFIM, one expects to observe an increase of the temperature of the critical point as well as reduction of the bias $\epsilon^{\ast}(T)$ required for coexistence of the liquid and the glass states.

For stronger disorder, the RFIM illustrates two possibilities for the response to the bias $s(>0)$, recall Sec.~\ref{sec:low_act_reg}.  For Gaussian disorder, the states of minimal activity are also free energy minima, so the bias does not generate new singularities, and the system exhibits a smooth crossover in activity, as $s$ increases.  Assuming $\mu>0$, this is accompanied by a crossover in the magnetisation, corresponding to an increasing overlap in the glassy case.  For bimodal disorder, Fig.~\ref{fig:3} shows that the state of minimal activity may differ from the equilibrium state (of minimal free energy).  In this case, increasing $s$ can lead to a first-order phase transition, even if there is no transition for the unbiased case $s=0$.  In the glassy context, this might result in a situation where the thermodynamic transition (induced by $\epsilon$) is absent, but the dynamical transition (induced by $s$) still survives.  Since we have seen that the dynamical phase behavior of the RFIM depends on the disorder distribution in this regime, it is not possible to make firm predictions for the corresponding glassy case, especially given that dependence of the mobility on the order parameter is also unknown at this stage.  Overall, our results reinforce the idea that the interplay of static and dynamic phase transitions can lead to a rich structure~\cite{jack2010large,Turner2015,jack2016phase}, even in simple mean-field models~\cite{van2010second,Jack2010rom,GuiothJack2020}.

%
  
\section{Conclusion}
\label{sec:conc}

We briefly summarise the main lessons that are available from this work.  

For the mean-field RFIM, the response to the bias $s$ can be analysed via the dynamical Landau free energy $\phi$.  The presence of disorder means that this quantity has a functional dependence on $\cal M$, which specifies the average magnetisation of the spins with a given value of the (random) magnetic field.  However, for discrete bimodal disorder (which is the simplest case), the function $\mdis$ is fully specified by just two numbers, so minimisation of the dynamical free energy is a numerically simple task, which proceeds similarly to minimisation of the equilibrium free energy (as described in Sec.~\ref{sec:eq-phase}).

As discussed in Sec.~\ref{sec:asymp-dyn}, there are some limits where minimisation of the dynamical free energy has a simple physical interpretation.  If $|s|$ is very large, one should maximise or minimise the activity, according to the sign of $s$.  In the RFIM, minimal activity states tend to be similar to equilibrium phases, while maximal activity states can be more complex, including for example states with magnetisation opposite to the applied field, similar to~\cite{van2010second,GuiothJack2020}.  If $|s|$ is small, then the dynamical free energy tends to be minimised by extrema of the equilibrium free energy; this can lead to dynamical phase transitions between its global minimum (which is dominant at $s=0$) and other extrema (either local minima or saddles, which may dominate for $s\neq 0$).  This effect is also observed in other similar models~\cite{Jack2010rom,GuiothJack2020}.

For the analogy between the RFIM and the RFOT theory of glasses, we have argued that differences in dynamics of liquid and glass phases can be partially captured by considering an RFIM with a mobility that does not respect the spin-flip symmetry of its equilibrium free energy.  This leads to the qualitative picture shown in Fig.~\ref{fig:9}, where increasing (positive) $s$ favours a glassy (high-overlap) dynamical phase, which is (qualitatively) consistent with~\cite{hedges2009dynamic}.
Given that the connection between the RFOT and RFIM theories remains somewhat abstract and there is no explicit connection between the dynamics of these models~\cite{biroli2014random,biroli2018random1,biroli2018random2,franz2013glassy,franz2013universality}, this theoretical picture leaves several open questions, the most serious of which is whether (and how) the trajectory ensemble of a glassy system can be connected in a precise way to the analogous ensembles of a model similar to the RFIM.  The consistency of  Fig.~\ref{fig:9} with dynamical properties of glassy systems~\cite{hedges2009dynamic} indicates that such a connection may be possible, but a detailed examination of this issue is (obviously) beyond the scope of this work.

\section*{Acknowledgements}

J.G. thanks the Royal Society for support through grant RP17002.
This work was funded in part by the European Research Council under the EU Horizon 2020 Programme,
grant number 740269.

\appendix

\section{Equilibrium Landau free energy of the mean-field RFIM}
\label{app:eq_landau_free_en}

We derive the equilibrium Landau free energy $\bar{f}$ as given in~\eqref{eq:49}, which is generic for any distribution $\pdis(h)$ of the (independent) random fields.
The mean-field RFIM has been studied in equilibrium in many works before \cite{schneider1977random,luttinger1976exactly,aharony1978tricritical,krapivsky2010kinetic}.  However, a notable feature of the current approach is that it does not use the replica method (as usually considered for the study of disordered systems).

For a fixed realisation of the quenched disorder $\bm{h}$, one can decompose the system into several subsystems, such that all spins of one subsystem share the same value of the random field $h_{i}$.
(If the random field has a continuous distribution then one should imagine grouping together spins with $h_i \in [h,h+dh]$.)
To this end, we define empirical distributions
\begin{eqnarray}
  \label{eq:3}
    \rho(h) = \frac{1}{N}\sum_{i=1}^{N} \delta(h-h_{i}) 
    \nonumber \\
     \varsigma(h)
    = \frac{1}{N} \sum_{i=1}^{N} \sigma_{i} \delta(h - h_{i}) \, .
\end{eqnarray}
For example, $N \int_{h_{1}}^{h_{2}}\! \rho(h)\, \mathrm{d}h$ is the number of sites $i\in \{1, \dots{}, N\}$ for which $h_{i} \in [h_{1}, h_{2}]$.
The total magnetisation and the overlap are then $ m = \int \varsigma(h)\, \mathrm{d}h $
and $\tau = \int h\varsigma(h)\, \mathrm{d}h $.
[Unless otherwise stated, integrals over $h$ are assumed to have limits $(-\infty,\infty)$.]
Using these formulae, the energy~\eqref{eq:1b} is completely specified by the empirical distribution $\varsigma(h)$. 

Since the random fields on each site are independent, one has by a law of large numbers that for $N\to\infty$ then $\rho(h) \to p_{\rm dis}(h)$ with probability one.  Moreover, the behavior in equilibrium states is that spins with similar random fields behave similarly, which implies $\varsigma(h) \to p_{\rm dis}(h) \mm(h)$ in the same limit, where $\mm(h)$ is the average magnetisation of spins with field $h$.  [Note: the convergence of $\rho\to\pdis$ relies only on the distribution of the disorder, but $\varsigma(h) \to p_{\rm dis}(h) \mm(h)$ is a statement on the distribution of the spins, which applies to typical configurations, and relies on the mean-field structure of the model.]

We now define a Landau free energy as a functional of $\mm$, which amounts to taking the log of a restricted canonical partition function, with the constraint that $\varsigma(h) \to \pdis(h) {\cal M}(h)$ as $N\to\infty$.  As usual, minimizing this Landau free energy over ${\cal M}$ will recover the equilibrium free energy, because the unrestricted partition function is dominated by configurations with $\varsigma$ close to the minimizer ${\cal M}^*$.  
The energy $E$ has already been expressed in terms of $\varsigma$, so the energy density ${\cal E}=E/N$ converges for large $N$ to
\begin{eqnarray}
\fl\quad\quad
{\cal E}[\mm] =
- \left(\int \pdis(h) \mm(h) \mathrm{d}h \right)^{2} - B \int \pdis(h) \mm(h) \mathrm{d}h - \eta \int h \pdis(h) \mm(h) \mathrm{d}h 
\nonumber\\
\end{eqnarray}
This quantity does not depend on the specific of the realisation of the disorder -- the assumption on $\varsigma$ implies that the energy is self-averaging.
It remains to characterise the entropy, which can be computed in terms of the number of spin configurations that are compatible with the distribution $\mm$, as $N\to\infty$.
The resulting entropy per spin is
${\cal S}[\mm|\bm{h}] = \int  \rho(h) S(\mm(h)) \, \mathrm{d}h $ where
\begin{equation}
  \label{eq:entropy}
  S(\mm) = - \left( \frac{1+\mm}{2} \ln \frac{1+\mm}{2} + \frac{1-\mm}{2}\ln\frac{1-\mm}{2} \right) \; .
\end{equation}
is the familiar entropy per spin for Ising spins with a constrained magnetisation.
Using again that $\rho\to\pdis$ at large $N$, the entropy is also self-averaging, and one arrives at an intensive Landau free energy
\begin{equation}
  \label{eq:38}
      \bar{f}[\mm] = {\cal E}[\mm] - \beta^{-1} \int \pdis(h) S(\mm(h)) \mathrm{d}h \; .
\end{equation}
As noted above, minimizing $\bar{f}$ over the distribution $\mm$ yields the equilibrium free energy $\bar{f}_{\rm eq}$ from (\ref{equ:bar-f-eq}).

\section{Dynamical Landau free energy}
\label{app:dyn_free_energy}

We evaluate the bound (\ref{eq:varppl}) using the ansatz (\ref{equ:pi_s_mf}), in order to obtain (\ref{eq:21}).  We start with the term
\begin{equation}
\sum_{\bm{\sigma}} r(\bm{\sigma}) \pi_s(\bm{\sigma}) = \sum_i \sum_{\bm{\sigma}}   w_1(\sigma_i,m,h_i) \pi_s(\bm{\sigma})
\label{equ:r-step1}
\end{equation}
where we used that the dynamics takes place by single spin flips, and $w_1$ is given in  (\ref{equ:w1}).  Since $\sigma_i=\pm 1$, one has for any function $f$ that 
\begin{equation}
f(\sigma_i)=\frac12 \left[ (1+\sigma_i) f(1) + (1-\sigma_i)f(-1)\right] \;.
\end{equation}
  Hence, recalling that $\pi_s$ depends on $\bm{h}$ but the individual spins are independent, (\ref{equ:pi_s_mf}) yields
\begin{equation}
\sum_{\bm{\sigma}}  f(\sigma_i) \pi_s(\bm{\sigma}) = \frac12 \left[ f(1)+f(-1) \right] + \frac{ \mm(h_i) }{2} \left[ f(1)-f(-1) \right] 
\end{equation}
Using this result with (\ref{equ:r-step1}) and also (\ref{eq:3}) yields 
\begin{equation}
\frac{1}{N}  \sum_{\bm{\sigma}} r(\bm{\sigma}) \pi_s(\bm{\sigma}) = \int \rho(h) r({\cal M}(h),h) \dst h + O(1/N)
\label{equ:r-step2}
\end{equation}
where the local escape rate $r$ given by (\ref{eq:14}); we suppress terms of order $N^{-1}$ that arise from the  correction at the same order in (\ref{eq:9}), since these will be irrelevant when we take $N\to\infty$ below.

We denote the second term on the right hand side of (\ref{eq:varppl}) as
\begin{equation}
F_2[w,\pi] = e^{-s} \sum_{\bm{\sigma}} \sum_{\bm{\sigma}'(\neq\bm{\sigma})} \sqrt{w(\bm{\sigma}|\bm{\sigma}')w(\bm{\sigma}'|\bm{\sigma})
  \pi(\bm{\sigma})\pi(\bm{\sigma}')}
\end{equation}
Using again (\ref{equ:pi_s_mf}) and that the dynamics takes place by single spin flips,
\begin{equation}
\fl \qquad F_2[w,\pi_s]  = e^{-s} \sum_i \sum_{\bm{\sigma}} \sqrt{ w_1(\sigma_i,m,h_i) w_1(-\sigma_i,m,h_i) } \pi_s(\bm{\sigma}) \sqrt{\frac{1-\sigma_i\mm(h_i)}{1+\sigma_i\mm(h_i)} }
\end{equation}
Repeating the same steps as before, one obtains a result analogous to (\ref{equ:r-step2}):
\begin{equation}
\frac{1}{N} F_2[w,\pi_s]  = e^{-s} \int 2 \rho(h) a(\mm(h),h) \dst  h + O(1/N)
\label{equ:F-step2}
\end{equation}
with the local mobility $a$ given by (\ref{eq:13}).

Finally, use (\ref{eq:small-psi}) with (\ref{eq:varppl}) to obtain
\begin{equation}
\psi(s) = -
\lim_{N\to\infty}\left\{  \min_{\pi} \frac{1}{N} 
\left[ \sum_{\bm{\sigma}} r(\bm{\sigma}) \pi(\bm{\sigma}) - F_2[w,\pi] \right] \right\}
\end{equation}
Assuming that the minimum is attained for some $\pi=\pi_s$ consistent with (\ref{equ:pi_s_mf}), one uses this
result with (\ref{equ:r-step2},\ref{equ:F-step2}) and the fact that $\rho(h)\to\pdis(h)$ as $N\to\infty$.  This yields (\ref{eq:21}), as required.

\section{Maxima of the global mobility for bimodal disorder}
\label{app:max_mob}

 For $s\ll -1$, the dynamical phase behaviour is controlled by the state of maximal mobility.  To investigate this, we investigate the stability of the stationary point $\mdis = 0$ of the mobility $\bar{a}$ when increasing the temperature $\beta$ or the disorder strength $\eta$. 
 Depending on whether this point is a maximum or a saddle, one may observe dynamical phase coexistence at $B=0$, see Figs.~\ref{fig:6}[A3] and~\ref{fig:5}[A].

It was shown in Sec.~\ref{sec:high_act} that $\mdis =0$ is always a stationary point of the global mobility $\bar{a}$. 
Since, $h_{i}=\pm 1$ for bimodal disorder, the distribution $\mdis$ is fully specified by two numbers $\bm{m}=(m_{+},m_{-})$, as $\mathcal{M}(h) = m_{+}\delta_{1}(h) + m_{-}\delta_{-1}(h)$.

A taylor expansion of $\bar{a}$ around $\bm{m}=0$, up to order $2$, leads to the Hessian
\begin{equation}
  \label{eq:31}
  \fl
  \bm{\mathcal{H}}_{\bar{a}}(0,0) = \frac{-1}{\cosh(\beta \eta)}
  \left[
  \begin{array}{cc} 
    {1+2\beta^{2}(1-2\tanh(\beta \eta)^{2})} &     {2\beta^{2}(1-2\tanh(\beta \eta)^{2})} \\
    {2\beta^{2}(1-2\tanh(\beta \eta)^{2})}    &        {1+2\beta^{2}(1-2\tanh(\beta \eta)^{2})}
  \end{array} 
  \right]
  \, ,
\end{equation}
whose eigenvalues are
\begin{eqnarray}
  \label{eq:32}
  \lambda_{+} &= -\frac{4\beta^{2}(1-2\tanh(\beta \eta)^{2}) + 1}{2\cosh(\beta \eta)} \\
  \lambda_{-} &= - \frac{1}{2\cosh(\beta \eta)} \nonumber \, .
\end{eqnarray}

One finds numerically that for $\beta < \beta^{\ast}=0.5$, both eigenvalues are always negative for any $\eta>0$ and thus the zero-overlap disordered state is the maximum of the global mobility $2\bar{a}$.
However, for $\beta \geqslant \beta^{\ast}=0.5$, one finds a second order transition (at which $\lambda_{+}=0$) located at
\begin{equation}
  \label{eq:20}
  \eta_{\rm mob}^{\ast}(\beta) = \frac{1}{\beta}\, \mathrm{arctanh}\, \sqrt{ \frac{1}{2}\left( 1 + {(2\beta)}^{-1}\right)} \, .
\end{equation}
One notes that $\eta_{\rm mob}^{\ast} \to +\infty$ for $\beta \to \beta^{\ast}=0.5$.
In the region of the parameter space where $\eta> \eta_{\rm mob}^{\ast}(\beta)$ and $\beta> \beta^{\ast}=0.5$, $(m,q)=(0,0)$ becomes a saddle-point and is no more a global maximum. Instead, two new magnetised states appear as symmetric maxima of the mobility: the symmetry between the two sub-systems $\Lambda_{+}$ and $\Lambda_{-}$ is spontaneously broken. As a consequence, the mobility of the spins in each sub-system is different ($a_{+}>a_{-}$ or $a_{-}<a_{+}$). This can be understood with the same argument already made below Eq.~\eqref{eq:28} (see also \cite{GuiothJack2020}): for large $\eta$, the state $\bm{m}=(0,0)$ looses its stability and the system is promoting states with lower entropic terms $\sqrt{1-\mathcal{M}(h)^{2}}$ but with one of the local field $\gamma_{\pm}=2m \pm \eta$ becoming smaller. A similar effect occurs in the activity-biased Ising model~\cite{GuiothJack2020}.


\section*{Bibliography}

\bibliographystyle{unsrt}
\bibliography{refs_rfim_activity}

\end{document}